\documentclass[usenatbib,fleqn,a4paper]{mn2e}

\usepackage[totalwidth=504pt, totalheight=690pt]{geometry}
\usepackage{amsmath}
\usepackage{amssymb}
\usepackage{graphicx}
\usepackage{mathptmx}
\usepackage{rotating}
\usepackage[T1]{fontenc}
\usepackage{aecompl}

\usepackage[colorlinks=true, citecolor=blue, linkcolor=blue, urlcolor=blue, breaklinks=true]{hyperref}
\hypersetup{pdftitle={Three-dimensional pure
  deflagration models with nucleosynthesis and synthetic observables
  for Type Ia supernovae}, pdfauthor={Michael Fink}}
\usepackage{microtype}
\usepackage{etoolbox}

\makeatletter

\patchcmd{\NAT@citex}
  {\@citea\NAT@hyper@{%
     \NAT@nmfmt{\NAT@nm}%
     \hyper@natlinkbreak{\NAT@aysep\NAT@spacechar}{\@citeb\@extra@b@citeb}%
     \NAT@date}}
  {\@citea\NAT@nmfmt{\NAT@nm}%
   \NAT@aysep\NAT@spacechar\NAT@hyper@{\NAT@date}}{}{}

\patchcmd{\NAT@citex}
  {\@citea\NAT@hyper@{%
     \NAT@nmfmt{\NAT@nm}%
     \hyper@natlinkbreak{\NAT@spacechar\NAT@@open\if*#1*\else#1\NAT@spacechar\fi}%
       {\@citeb\@extra@b@citeb}%
     \NAT@date}}
  {\@citea\NAT@nmfmt{\NAT@nm}%
   \NAT@spacechar\NAT@@open\if*#1*\else#1\NAT@spacechar\fi\NAT@hyper@{\NAT@date}}
  {}{}

\makeatother

\def\aj{AJ}%
\def\araa{ARA\&A}%
\def\apj{ApJ}%
\def\apjl{ApJ}%
\def\aap{A\&A}%
\def\mnras{MNRAS}%
\def\pasp{PASP}%
\def\nat{Nature}%
\newcommand{\gcc}{\ensuremath{\text{g} \, \text{cm}^{-3}}}
\newcommand{\kms}{\ensuremath{\text{km} \, \text{s}^{-1}}}
\newcommand{\msol}{\ensuremath{\text{M}_\odot}}
\newcommand{\mch}{\ensuremath{\text{M}_\text{Ch}}}
\newcommand{\nuc}[2]{\ensuremath{^{#1}\text{#2}}}

\title[3D pure deflagration models for SNe~Ia]{Three-dimensional pure
  deflagration models with nucleosynthesis and synthetic observables
  for Type~Ia supernovae}

\author[M.~Fink et al.]{Michael~Fink$^{1}$%
  \thanks{E-mail:
    \href{mailto:mfink@astro.uni-wuerzburg.de}{mfink@astro.uni-wuerzburg.de}}
  Markus~Kromer$^{2,3}$,
  Ivo~R.~Seitenzahl$^{1,2}$,\newauthor
  Franco~Ciaraldi-Schoolmann$^{2}$,
  Friedrich~K.~R\"opke$^{1}$,
  Stuart~A.~Sim$^{4}$,\newauthor
  R\"udiger~Pakmor$^{5}$,
  Ashley~J.~Ruiter$^{2}$ and
  Wolfgang~Hillebrandt$^{2}$\\ \\
  $^{1}$Institut f\"ur Theoretische Physik und Astrophysik,
  Universit\"at W\"urzburg, Emil-Fischer-Stra{\ss}e 31, D-97074
  W\"urzburg, Germany\\
  $^{2}$Max-Planck-Institut f\"ur Astrophysik,
  Karl-Schwarzschild-Stra{\ss}e 1, D-85748 Garching, Germany\\
  $^{3}$The Oskar Klein Centre, Stockholm University,
  AlbaNova, SE-106 91 Stockholm, Sweden\\
  $^{4}$Astrophysics Research Centre, School of Mathematics and
  Physics, Queen's University Belfast, Belfast BT7 1NN, UK\\
  $^{5}$Heidelberger Institut f\"{u}r Theoretische Studien,
  Schloss-Wolfsbrunnenweg 35, D-69118 Heidelberg, Germany}

\defcitealias{seitenzahl2013a}{S13}

\begin{document}

\date{Accepted 2013 November 29.  Received 2013 November 29; in
  original from 2013 August 14}
\volume{438} \pagerange{1762--1783} \pubyear{2014}
\maketitle

\begin{abstract}
  We investigate whether pure deflagration models of
  Chandrasekhar-mass carbon--oxygen white dwarf stars can account for
  one or more subclass of the observed population of Type~Ia
  supernova (SN~Ia) explosions.
  We compute a set of 3D full-star hydrodynamic explosion models, in
  which the deflagration strength is parametrized using the multispot
  ignition approach.  For each model, we calculate detailed
  nucleosynthesis yields in a post-processing step with a 384 nuclide
  nuclear network.  We also compute synthetic observables with our 3D
  Monte Carlo radiative transfer code for comparison with
  observations.
  For weak and intermediate deflagration strengths (energy release
  $E_\text{nuc} \lesssim 1.1 \times 10^{51}$~erg), we find that the
  explosion leaves behind a bound remnant enriched with $3$ to $10$
  per cent (by mass) of deflagration ashes.  However, we do not obtain
  the large kick velocities recently reported in the literature.  We
  find that weak deflagrations with $E_\text{nuc} \sim 0.5 \times
  10^{51}$~erg fit well both the light curves and spectra of
  2002cx-like SNe~Ia, and models with even lower explosion energies
  could explain some of the fainter members of this subclass.  By
  comparing our synthetic observables with the properties of SNe~Ia,
  we can exclude the brightest, most vigorously ignited models as
  candidates for any observed class of SN~Ia: their $B - V$ colours
  deviate significantly from both normal and 2002cx-like SNe~Ia and
  they are too bright to be candidates for other subclasses.
\end{abstract}


\section{Introduction}
\label{sec:int}

It is generally accepted that Type~Ia supernovae (SNe~Ia) originate
from a thermonuclear explosion of a carbon--oxygen white dwarf (CO WD)
in an interacting binary system \citep[see][for a current review on
constraints on the progenitor system]{wang2012b}.  However, despite
decades of research, SNe~Ia remain incompletely understood: neither
the exact nature of their progenitor systems nor the burning mode in
which the explosions proceed are clearly identified.

A well-established progenitor channel is the single-degenerate
Chandrasekhar-mass (\mch) scenario, in which the companion star to the
WD is either a main-sequence or evolved non-degenerate star.  In this
scenario, the explosion is triggered when the WD approaches \mch\ due
to accretion from its companion.  Recent observations of time-varying
Na features in the spectra of some SNe~Ia point towards a
single-degenerate origin \citep{patat2007a, sternberg2011a,
  dilday2012a}: the Na features may be interpreted as signatures of
nova shells that were expelled from the WD during the accretion phase
(see, however, \citealt*{shen2013a} and \citealt{soker2013a} for
alternative explanations).

Within the single-degenerate scenario, several explosion mechanisms
are possible (see \citealt{hillebrandt2000a} for a review).  It is
known that the explosion has to start as a subsonic deflagration.
Unlike for a prompt detonation, a deflagration allows parts of the WD
to expand significantly before being burnt.  Thus, a sufficient amount
of intermediate-mass elements (IMEs), such as Si, can be produced, as
needed to explain observed SN~Ia spectra.  To be consistent with
normal SNe~Ia, the explosion likely has to turn into a detonation at
later burning stages (so-called delayed detonation models) to
reproduce the abundance stratification inferred from spectral
evolution \citep[e.g.][]{stehle2005a} and also to reach observed
brightnesses/explosion energies.  There are different mechanisms that
could trigger such a secondary detonation.  In the
deflagration-to-detonation transition scenario (DDT;
\citealt{blinnikov1986a, blinnikov1987a, khokhlov1991a}), the
deflagration is supposed to spontaneously turn into a detonation at
late burning stages after the flame has entered the distributed
burning regime.  Other scenarios initiate a detonation in regions that
are compressed due to fallback of hot deflagration ashes that stay
gravitationally bound [e.g., so-called gravitationally confined
detonation (GCD) scenarios; cf.\ \citealt*{plewa2004a}; see
Section~\ref{sec:puls_secdet} for details].

Despite substantial modelling efforts \citep*{ciaraldi2013a}, DDTs are
still not well understood, and \citet{niemeyer1999a} even argue that
they may not occur at all.  The success of the GCD-like scenarios
depends on the amount of energy released in the deflagration and thus
on the details of the initial flame geometry, which are not well
constrained yet.  Therefore, some explosions likely occur as pure
deflagrations.  This scenario has been suggested as potential
explanation of the peculiar subclass of 2002cx-like SNe~Ia
\citep{branch2004a, jha2006b, phillips2007a}.

Apart from the single-degenerate \mch\ scenario, there are alternative
models for SNe~Ia \citep[see e.g.][]{hillebrandt2013a}.  Both violent
WD--WD mergers \citep{pakmor2013a, ruiter2013a} and sub-Chandrasekhar
double detonations \citep{fink2010a, kromer2010a, woosley2011b,
  moll2013a} may account for normal and subluminous types of SNe~Ia.
However, the mixed abundance patterns of 2002cx-like SNe~Ia (inferred
from their peculiar spectra; \citealt{jha2006b, phillips2007a}) seem
to be inconsistent with those scenarios, in which explosive burning
takes place as detonation (resulting in layered abundance patterns).
Thus, \mch\ pure deflagrations may still be required in order to
explain the full range of subclasses of SNe~Ia.

Pure deflagrations have been studied extensively in numerical
simulations (cf.\ \citealt*{reinecke2002d}; \citealt{gamezo2003a,
  garcia2005a, roepke2006a, roepke2006b, roepke2007c, jordan2012b,
  long2013a, ma2013a} for recent multidimensional studies).  To
investigate their potential contribution to the observed sample, we
carry out an extensive study of pure deflagrations in 3D
Chandrasekhar-mass models for a very wide range of explosion
strengths, including detailed nucleosynthetic post-processing and for
the first time multidimensional radiative transfer for pure
deflagrations (only for one of the models of this study, N5def,
synthetic observables have already been published in a companion
paper, \citealt{kromer2013a}).  The computed synthetic observables can
be directly compared to real observations, which allows us both to
constrain the assumed range of initial flame geometries and to
investigate which classes of observed SNe~Ia might be explained with
pure deflagration models.

This study is organized as follows.  In Section~\ref{sec:expl_scen}, we
present our WD models and the initial flame geometries that are used
to initiate deflagration burning in our simulations.  Our numerical
methods are summarized in Section~\ref{sec:num_methods}.  Then, we
present the results of our hydrodynamic simulations
(Section~\ref{sec:hd_expl_mod}) and our nucleosynthesis calculations
(Section~\ref{sec:nucl}).  In Section~\ref{sec:synth_obs}, we show the
outcomes of our radiative transfer calculations and compare our
synthetic light curves and spectra with observations of real
supernovae (SNe).  Finally, we summarize our results in
Section~\ref{sec:disc}.

\section{Explosion scenario}
\label{sec:expl_scen}

In this work, we perform 3D hydrodynamic calculations to simulate the
explosion of a full WD star.  Our simulations are closely related to
those in \citet[hereafter referred to as S13]{seitenzahl2013a}: we use
the same initial WD models and identical setups for the deflagrations.
In contrast to \citetalias{seitenzahl2013a}, however, we assume that
no delayed detonation occurs and consider the corresponding pure
deflagration models.  In the following, we summarize the main
parameters of our models (see \citetalias{seitenzahl2013a} for further
details).

In our standard initial models, we adopt \mch\ CO WDs in hydrostatic
equilibrium, each with a central density of $\rho_\text{c} = 2.9
\times 10^9~\gcc$ and a constant temperature $T = 5 \times
10^5~\text{K}$.  To test the influence of the central density, two
models were set up with different central densities of $1.0$ and $5.5
\times 10^9~\gcc$ (N100Ldef and N100Hdef).  All models assume a
homogeneous initial composition with mass fractions $X_{^{12}\text{C}}
= 0.475$, $X_{^{16}\text{O}} = 0.50$ and $X_{^{22}\text{Ne}} = 0.025$,
which approximately correspond to a solar metallicity of the zero-age
main-sequence progenitor star that evolved to the WD\@.

We start our simulations at the onset of the thermonuclear explosion
and do not simulate the previous evolution.  As argued in
\citetalias{seitenzahl2013a}, the conditions at the deflagration
ignition are not finally settled.  Thus, in this study (as in
\citetalias{seitenzahl2013a}) we use the pragmatic approach of
igniting the deflagration in multiple spherical ignition spots that
start burning simultaneously (so-called multispot ignition) and treat
the ignition geometry as a free parameter.  In the case of a large
number of ignition spots, the flame is effectively always ignited
centrally.  Such central ignitions with many seeds for instabilities
have been found to be the only way of allowing pure deflagration
explosions that reach the brightness of (faint) normal SNe~Ia
(\citealt{roepke2006a}; but, see also \citealt{ma2013a}, who find even
larger \nuc{56}{Ni} masses).  In the case of a small number of
ignition spots, the outermost ignition spot tends to dominate the
burning, which leads to a one-sided deflagration flame in most of our
models.  This is consistent with the results of recent pre-ignition
simulations that predict bipolar flows through the centre in the
simmering phase and a one-sided off-centre ignition
\citep{nonaka2012a}.  Rotation, on the other hand, could suppress such
bipolar flows and render a central ignition more likely.

\begin{figure*}
  \begin{center}
    \includegraphics[width=\textwidth]{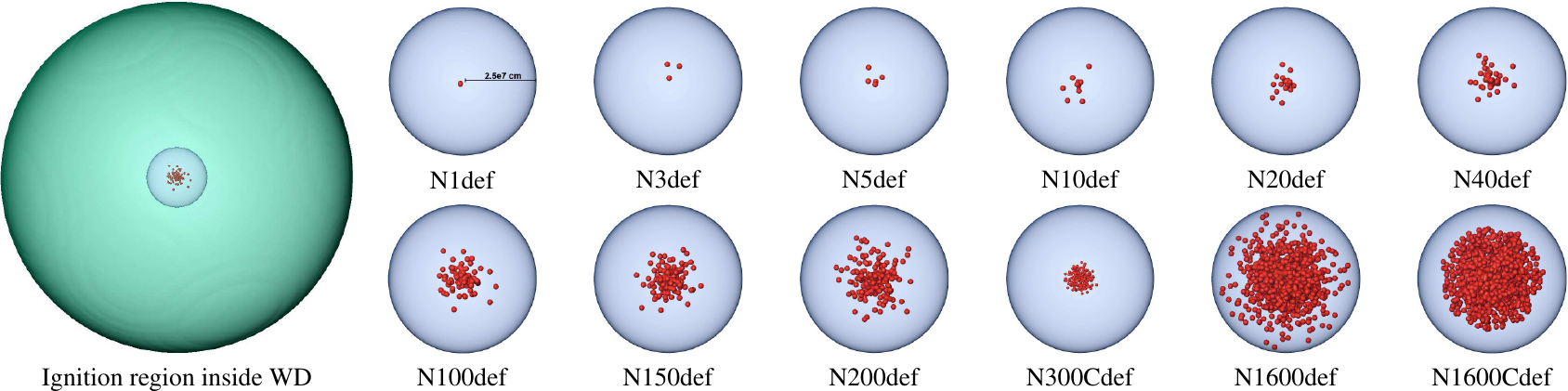}
    \caption{Ignition geometries of our deflagration models.  Shown
      are the ignition kernels (red spheres) and a transparent blue
      contour where the distance to the centre is 250~km.  The maximum
      radii $r_\text{max}$ of the ignition kernel distributions are
      36, 67, 81, 91, 91, 137, 130, 146, 172, 65, 250 and 180~km,
      respectively.  The leftmost panel shows the position of the
      ignition regions within the WD\@.}
    \label{fig:ign}
  \end{center}
\end{figure*}

With our multiple ignition spot parametrization we cannot cover all
potential initial flame geometries, but we can access a wide range of
explosion strengths in a numerically well-behaved manner (see
\citetalias{seitenzahl2013a}, for details).  We assume ignitions in
$N_\text{k} = 1$, 3, 5, 10, 20, 40, 100, 150, 200, 300 and 1600 (two
realizations) sparks placed randomly around the centre of the WD (see
Fig.~\ref{fig:ign}).\footnote{We stress that values of $N_\text{k}$ as
  high as 1600 are merely used to parametrize a central deflagration
  that is already strongly developed at early explosion stages.
  Physically, deflagration ignition is not likely to occur
  simultaneously in such a large number of spots.}
Two models (N300Cdef and N1600Cdef) are constructed to have a very
dense arrangement of the ignition kernels (with low $r_\text{max}$)
and are referred to as `compact' models with an additional `C' in the
model name.  All other models (with standard initial WD) are depicted
`standard' models.  The models with alternative central densities
(N100Ldef and N100Hdef) use exactly the same spatial positions of the
ignition sparks as the N100def model.  The radius $R_\text{k}$ of the
spherical ignition kernels is always 10~km.  Only model N300C, which
has the most compact configuration with large $N_\text{k}$, uses a
lower value of $R_\text{k} = 5~\text{km}$.

\section{Numerical methods}
\label{sec:num_methods}

In our simulations, we use the same numerical methods as in
\citetalias{seitenzahl2013a} and \citet{sim2013a}.  Therefore, we only
give a brief summary here (see above references for more details).

\subsection{Hydrodynamic simulation}
\label{sec:hd}

Our hydrodynamical simulations are carried out in three dimensions
with the finite volume hydrodynamics code \textsc{leafs}
\citep*{reinecke2002b} on a $512^3$-cells `hybrid' moving grid
\citep{roepke2005c}.  The flame evolution is tracked by an inner
Cartesian part of the grid (initial spatial resolution: 1.9~km); the
outer parts of the grid, which have exponentially growing cell sizes,
track the expansion of the explosion ejecta.  With this co-expanding
grid, the ejecta evolution is efficiently followed until $t = 100$~s
(by which time homologous expansion is a good approximation for most
models).  This advantage comes at the expense of a poor resolution of
the central bound remnants in some of our models (see
Section~\ref{sec:unbinding_wd}).  We emphasize, however, that our main
intention is to derive synthetic observables around peak luminosity.
Therefore, we are mainly interested in the ejected material that
causes emission at this epoch.

We treat the flame as an infinitesimally thin surface that separates
fuel and ashes.  The change of position of the flame is described with
a level set approach (\citealt{osher1988a};
\citealt*{smiljanovski1997a}; \citealt{reinecke1999a}).  Changes of
the chemical composition and the release of nuclear energy are
performed instantaneously behind the front.  We only use a reduced set
of five species in the hydrodynamic simulations: $^4$He, $^{12}$C,
$^{16}$O, and representative species for both IMEs and iron-group
elements (IGEs).\footnote{In our hydrodynamic simulations our initial
  WD composition is represented as $X_{^{12}\text{C}} = 0.5$ and
  $X_{^{16}\text{O}} = 0.5$ with an electron fraction of $Y_\text{e} =
  0.498\,86$, which corresponds to a $^{22}$Ne content of $2.5$ per
  cent by mass.}  The (reduced) composition of the ashes and the
energy release as a function of fuel density is taken from
pre-calculated tables (see Appendix~\ref{sec:deftab}).  The adjustment
of the nuclear statistical equilibrium (NSE) composition to the
changing thermodynamic background state \citep[see][]{seitenzahl2009a}
and electron captures is also taken into account.  As we cannot fully
resolve the turbulent structures that have an impact on the flame
surface and thus the burning speed, a subgrid-scale turbulence model
is applied to determine an effective burning velocity which is valid
on our grid scale (for details, see \citealt*{schmidt2006b};
\citealt{schmidt2006c}).  For the flame--flow coupling, we use the
`passive implementation' described in \citet{reinecke1999a}.

Self-gravity is included using the monopole approximation for the
gravitational potential.  To test the influence of asymmetries in the
gravitational field on our simulation results, two of our models were
re-calculated using a fast Fourier transformation-based algorithm for
solving the Poisson equation.  These calculations are performed on a
$512^3$-cells uniform grid tracking the WD expansion and are marked
with a suffix `FFT' in the model names.

\subsection{Nucleosynthesis post-processing}
\label{sec:pp}

We determine the detailed nucleosynthetic yields of the ejecta in a
post-processing step after the hydrodynamic simulation.  To this end,
we solve a large nuclear network containing 384 species (ranging up to
$^{98}$Mo; see \citealt{travaglio2004a}) for the trajectories of
$10^6$ equal mass tracer particles, which are passively advected in
the hydrodynamic simulation.  We use an updated version of the RE\-AC\-LIB
reaction rate library \citep[][updated 2009]{rauscher2000a}.

\subsection{Radiative transfer}
\label{sec:rt}

For our radiative transfer calculations, the final ejecta density and
the final post-processing abundances of the tracer particles (which
have irregularly distributed coordinates) are mapped on a $200^3$
Cartesian grid in asymptotic velocity space with a
smoothed-particle-hydrodynamics-like algorithm that accurately
conserves the integrated yields \citep[for details,
see][]{kromer2010a}.  After further down sampling the data to a $50^3$
Cartesian grid (velocity resolution: $400$--$600~\kms$), we perform a
time-dependent 3D Monte Carlo radiative transfer simulation with the
\textsc{artis} code \citep{sim2007b, kromer2009a}.  For each model we
use $10^8$ photon packets and follow 111 logarithmically spaced time
steps between 2 and 120~d after explosion ignition.  We use the
`cd23\_gf-5' atomic data set of \citet{kromer2009a} expanded to
include ions up to \textsc{vii} for elements heavier than Ca but
lighter than Cu.  To reduce the computational costs, a grey
approximation is used in optically thick cells
\citep[cf.][]{kromer2009a} and for $t < 3$~d, local thermodynamic
equilibrium is assumed.

\section{Hydrodynamic explosion models}
\label{sec:hd_expl_mod}

\subsection{Flame propagation}
\label{sec:flame_prop}

\begin{figure*}
  \centering
  \includegraphics[width=14.8cm]{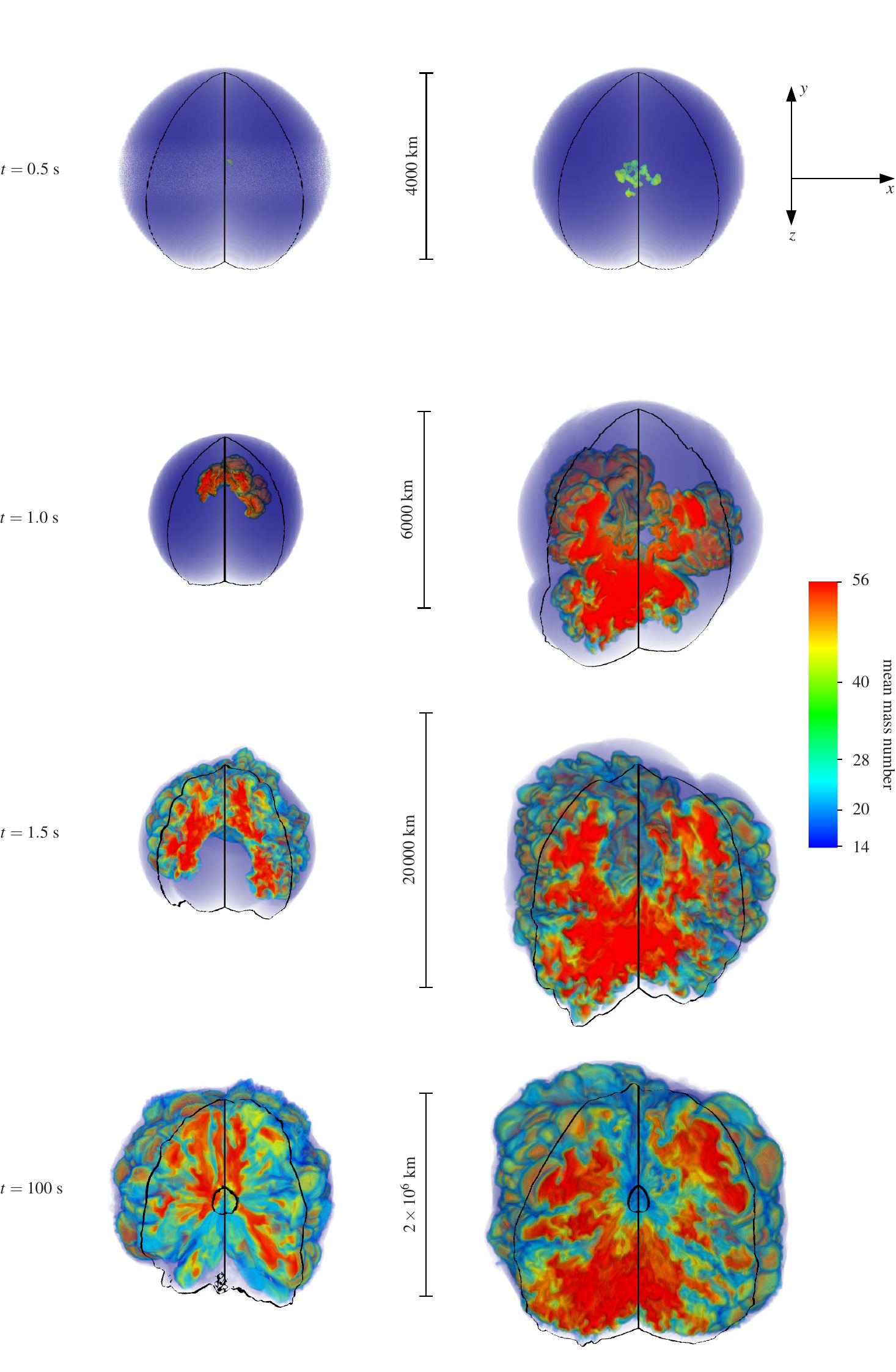}
  \caption{Explosion evolution for models N1def (left-hand column) and
    N20def (right-hand column).  Colour coded is the mean mass number
    calculated from the reduced set of species in the hydrodynamic
    simulation.  In the volume renderings a $90^{\circ}$ wedge is cut
    out in the front part of the ejecta.  The times after explosion
    initiation are from top to bottom: $t = 0.5$, $1.0$, $1.5$ and
    $100$~s (for each time the length-scale along the middle axis of
    the plots is given in the centre).  At $t = 100$~s, the innermost
    black contours mark the outer edges of the regions which do not
    become gravitationally unbound.}
  \label{fig:expl_evol}
\end{figure*}

\begin{figure*}
  \centering
  \includegraphics[width=16.8cm]{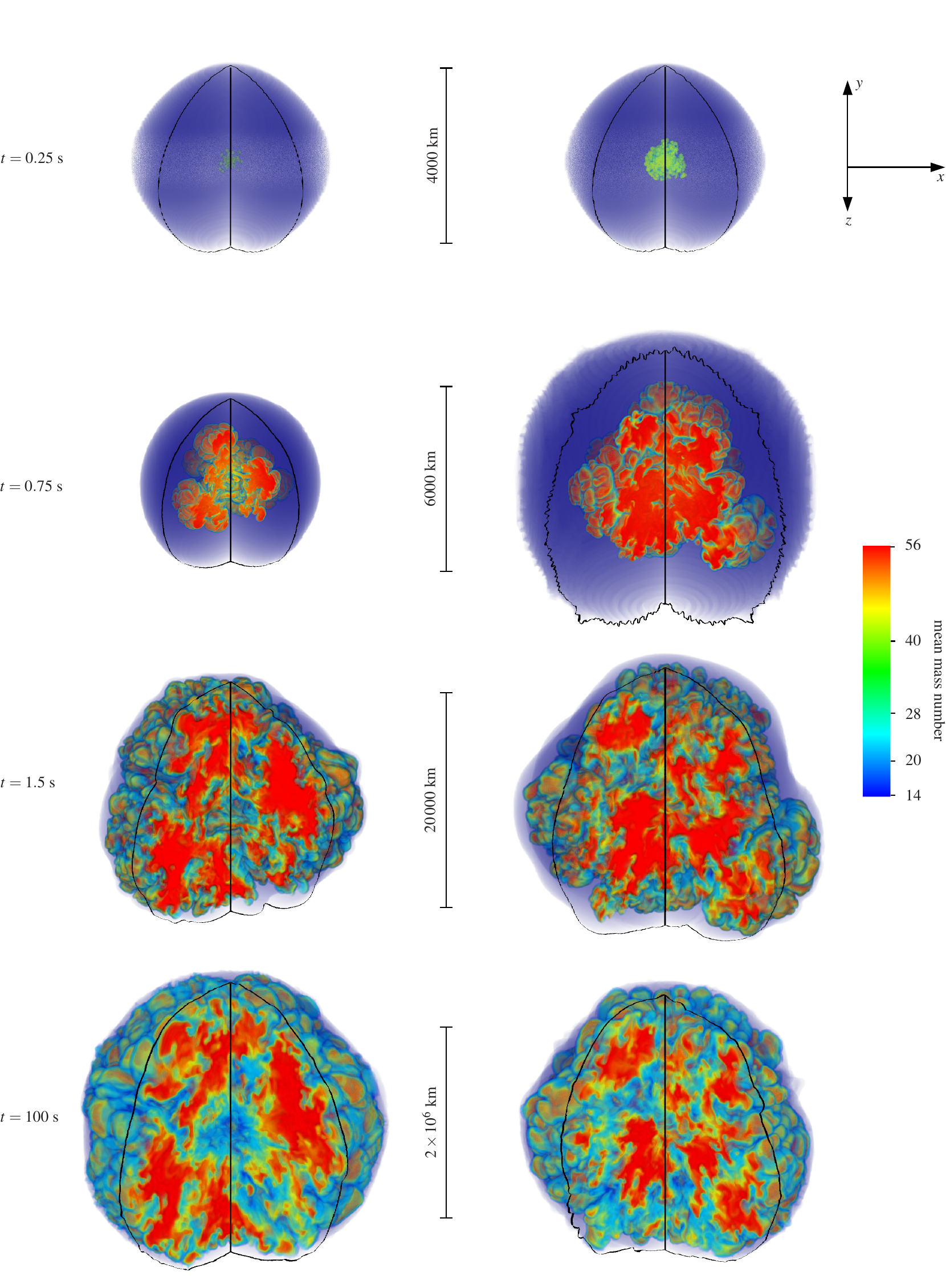}
  \caption{As Fig.~\ref{fig:expl_evol}, but showing models N150def
    (left-hand column) and N1600def (right-hand column).  The times
    after explosion initiation are from top to bottom: $t = 0.25$,
    $0.75$, $1.5$ and $100$~s.}
  \label{fig:expl_evol2}
\end{figure*}

As mentioned in Section~\ref{sec:expl_scen}, in all of our 14 models,
the burning takes place exclusively as subsonic deflagration and is
initiated in multiple ignition spots close to the WD's centre (see
Fig.~\ref{fig:ign}).  At the beginning of each simulation, the flame
propagates at the laminar deflagration speed.  Later, the burning is
accelerated by instabilities and the interaction with turbulence.
Since the growth of these instabilities is strongly influenced by the
deflagration ignition geometry, different models show very different
rates of flame spreading and acceleration.  In the following
discussion of our results, we will mainly focus on the four standard
models N1def, N20def, N150def and N1600def (see
Figs~\ref{fig:expl_evol} and \ref{fig:expl_evol2}),
which are representative for the whole sample and exemplify most of
its variety.

Model~N1def is a realization of the simplest case, the ignition of a
single spherical bubble at a location slightly off-centre (distance
from the origin $r = 36.5$~km and bubble radius $R = 10$~km).  Its
evolution is illustrated in the left-hand column of
Fig.~\ref{fig:expl_evol} and is a result of the buoyant rise of the
hot ashes in the initial burning bubble combined with the propagation
due to burning: initially the bubble rises without burning much
material (see snapshot at $t = 0.5$~s in the left-hand column of
Fig.~\ref{fig:expl_evol}), then it spreads laterally, burning a
relatively contiguous volume.  However, most of the high-density
material in the centre of the WD is left unburnt ($t = 1.0$~s).  The
buoyant rise of the bubble and the expansion of the star due to the
liberation of nuclear energy causes the flame to extinguish before it
is able to completely wrap around the unburnt core\footnote{The
  extinction density for C/O deflagrations is ${\sim} 5 \times
  10^6~\gcc$.} ($t = 1.5$~s).  Nevertheless, even after flame
extinction the hot ashes continue to expand and spread around the
unburnt central parts and finally cover almost the full solid angle
($t = 100$~s, when the ejecta have reached homologous expansion to a
good approximation).

Other models with a small number of ignition sparks ($N_\text{k} \le
20$) evolve in a manner similar to N1def.  The right-hand column of
Fig.~\ref{fig:expl_evol} shows N20def as an example: due to their
separation, most bubbles initially rise up individually without much
interaction (see snapshot at $t = 0.5$~s) and neighbouring burning
fronts only start to merge in the outer layers (see e.g. $t = 1.0$~s
and later).  Therefore, apart from the fact that more matter is burnt,
the final distribution of burning products is relatively similar to
that of N1def.  At $t = 1.5$~s, when burning ceases, N20def also shows
a Ni-rich outer layer that is wrapped around a central region that is
mostly unburnt.  However, the greater energy release due to the more
complete burning of model N20def compared to N1def also leads to
higher expansion velocities in the final ejecta (see $t = 100$~s in
Fig.~\ref{fig:expl_evol}).  In addition, model N20def has a
significantly different inner ejecta structure from N1def: despite
having a lower mass of unburnt C/O close to the centre, this material
occupies a much bigger volume than in N1def.  This behaviour is due to
differences in the transfer/redistribution of kinetic energy from the
ashes to the unburnt fuel (see Section~\ref{sec:vel_nucl} for further
discussion of the final ejecta structures) and is closely linked to
the occurrence and size of a bound remnant that may form for all
models with sparsely ignited deflagrations (see
Section~\ref{sec:unbinding_wd}).

\begin{figure}
  \centering
  \includegraphics[width=\columnwidth]{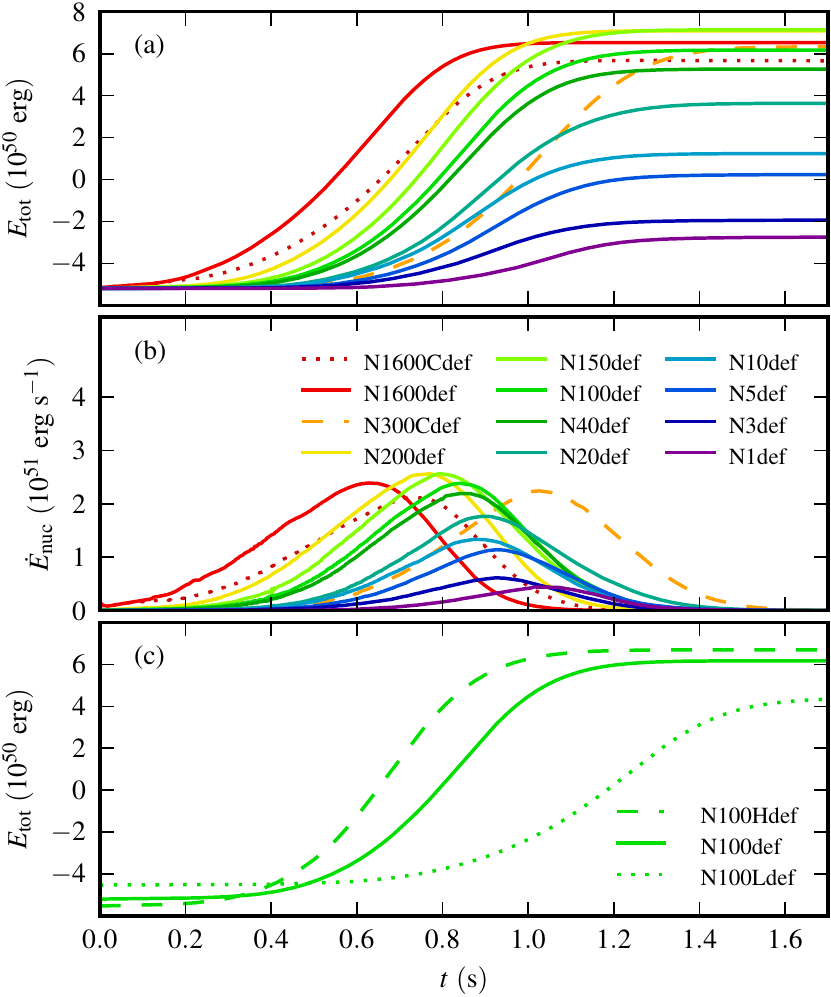}
  \caption{Temporal evolution of total energy (a) and rate of nuclear
    energy release (b) for all models with our standard initial WD
    model ($\rho_\text{c} = 2.9 \times 10^9~\gcc$) in the initial
    $1.7$~s.  The total energy evolution for the N100 deflagration
    models with different WD central densities is shown in panel~(c).}
  \label{fig:etot}
\end{figure}

In models with many ignition sparks ($N_\text{k} \gtrsim 100$, see
Fig.~\ref{fig:expl_evol2}) both burning and expansion take place much
more rapidly than in models with few ignition sparks.  The flame
evolution is no longer one sided, but large flame structures form in
all directions.  Due to its larger $N_\text{k}$, model N1600def
evolves faster than model N150def, but since it burns less mass, the
final ejecta velocities are lower (see Section~\ref{sec:expl_ene} and
Fig.~\ref{fig:etot}
for the explosion energetics).

Model N150def shows the typical ejecta structure of a strong
deflagration: a limited number of large rising plumes of burnt matter
(which often reach out to the highest velocities) and downdrafts of
unburnt fuel in between.  The centre is dominated by unburnt C/O
material.  Due to the huge number of ignition spots, model N1600def
contains less unburnt matter in the centre.  However, in a realistic
multidimensional treatment, deflagration burning is never
`volume filling' and always develops downdrafts of unburnt matter
between rising burning plumes.  Therefore, even in this most extreme
model, there are several unburnt fingers of C/O material that reach
down to the centre.

\subsection{Explosion energetics}
\label{sec:expl_ene}

\begin{figure}
  \centering
  \includegraphics[width=\columnwidth]{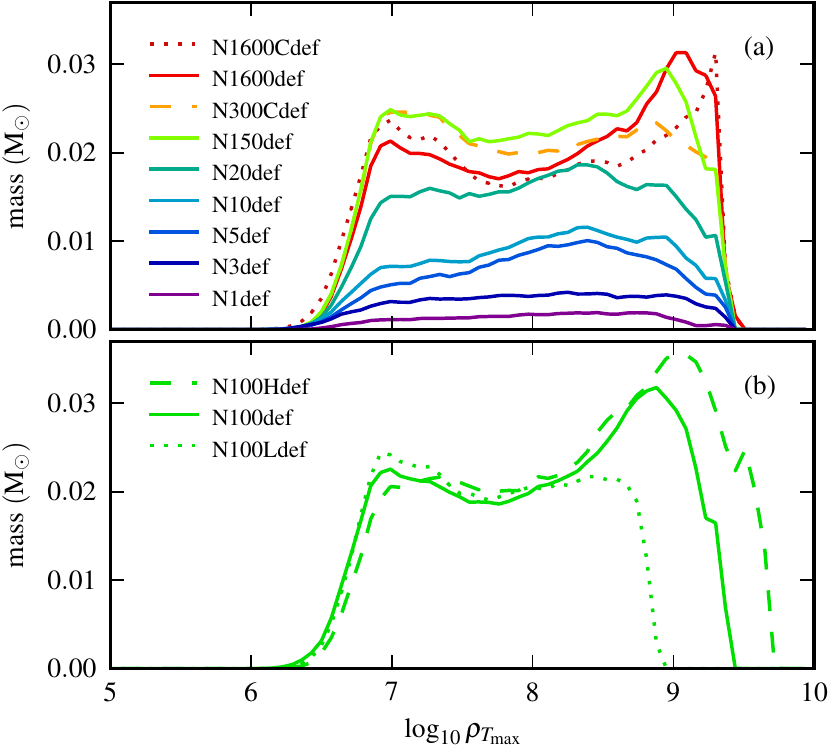}
  \caption{Distribution of ejected mass as a function of the density
    at which it was burnt ($\rho_{T_\text{max}}$ is the density at the
    maximum temperature experienced by a tracer particle): (a) for
    select models with $\rho_\text{c} = 2.9 \times 10^9~\gcc$ and
    (b)~for the N100 deflagration models with different central
    densities.}
  \label{fig:massrho}
\end{figure}

Our series of models produces a wide range of explosion energies.
Total energies ($E_\text{tot} = E_\text{int} + E_\text{kin} +
E_\text{grav}$) and nuclear energy releases ($E_\text{nuc}$) have been
calculated globally for the whole star at $t = 100$~s (see
Table~\ref{tab:energetics},
\begin{table*}
  \centering
  \begin{minipage}{146mm}
    \caption{Total masses (in solar masses) of the ejecta, the bound
      remnants and the main ejected nucleosynthesis products and
      explosion energies (in units of $10^{50}$~erg) of all models.}
    \label{tab:energetics}
    \begin{tabular}{@{}lccccccccccc@{}}
      \hline
      Model & $M_\text{ej}$ & $M_\text{b}$ & $M_{^{56}\text{Ni}}$ & $M_\text{IGE}$ & $M_\text{IME}$ &
      $\frac{M_{^{56}\text{Ni}}}{M_\text{IGE}}$ & $\frac{M_\text{IGE} + M_\text{IME}}{M_\text{ej}}$ &
      $E_\text{nuc}$ & $\frac{E_\text{nuc}}{|E_\text{bind}|}$ & $E_\text{tot}$ & $E_\text{kin,ej}$ \\
      \hline
      N1def   &$0.0843$& $1.32$ & $0.0345$& $0.0468$& $0.00893$& $0.74$ & $0.66$ & $1.74$ & $0.34$& $-0.580$ & $0.149$ \\
      N3def   & $0.195$& $1.21$ & $0.0730$& $0.106$ & $0.0257$ & $0.69$ & $0.68$ & $3.06$ & $0.59$& $-0.104$ & $0.439$ \\
      N5def   & $0.372$& $1.03$ & $0.158$ & $0.222$ & $0.0416$ & $0.71$ & $0.71$ & $4.90$ & $0.94$& $0.962$ & $1.35$ \\
      N10def  & $0.478$& $0.926$& $0.183$ & $0.267$ & $0.0581$ & $0.69$ & $0.68$ & $5.87$ & $1.1$ & $1.68$ & $1.95$ \\
      N20def  & $0.859$& $0.545$& $0.264$ & $0.394$ & $0.125$  & $0.67$ & $0.60$ & $8.36$ & $1.6$ & $3.68$ & $3.75$ \\
      N40def  & $1.21$ & $0.190$& $0.335$ & $0.509$ & $0.142$  & $0.66$ & $0.54$ & $10.2$ & $2.0$ & $5.26$ & $5.22$ \\
      N100Ldef& $1.23$ & $0.133$& $0.326$ & $0.413$ & $0.138$  & $0.79$ & $0.45$ & $8.79$ & $1.9$ & $4.36$ & $4.32$ \\
      N100def & $1.31$ & $0.090$& $0.355$ & $0.545$ & $0.147$  & $0.65$ & $0.53$ & $11.1$ & $2.1$ & $6.16$ & $6.11$ \\
      N100Hdef& $1.31$ & $0.102$& $0.329$ & $0.576$ & $0.179$  & $0.57$ & $0.58$ & $11.8$ & $2.1$ & $6.68$ & $6.63$ \\
      N150def & $1.40$ &($0.048$)&$0.378$ & $0.583$ & $0.164$  & $0.65$ & $0.53$ & $12.1$ & $2.3$ & $7.12$ & $6.98$ \\
      N200def & $1.40$ &($0.021$)&$0.371$ & $0.598$ & $0.146$  & $0.62$ & $0.53$ & $12.0$ & $2.3$ & $7.07$ & $6.95$ \\
      N300Cdef& $1.40$ &($0.027$)&$0.334$ & $0.526$ & $0.167$  & $0.63$ & $0.50$ & $11.2$ & $2.2$ & $6.31$ & $6.26$ \\
      N1600def& $1.40$ &($0.016$)&$0.340$ & $0.582$ & $0.132$  & $0.58$ & $0.51$ & $11.5$ & $2.2$ & $6.50$ & $6.38$ \\
      N1600Cdef&$1.40$ &($0.016$)&$0.315$ & $0.532$ & $0.136$  & $0.59$ & $0.48$ & $10.7$ & $2.1$ & $5.63$ & $5.50$ \\
      \hline
    \end{tabular}
    $M_\text{ej}$ and $M_\text{b}$ are the total masses of the ejecta
    and the bound material; $M_{^{56}\text{Ni}}$, $M_\text{IGE}$ and
    $M_\text{IME}$ are the ejected masses of $^{56}$Ni, IGEs and IMEs
    (as determined in the nucleosynthesis post-processing).  The small
    bound masses (values in brackets) of vigorously ignited models
    were neglected and considered as part of the ejecta.
    $E_\text{kin,ej}$ is the asymptotic kinetic energy of the ejecta,
    the other energy values are calculated for the whole WD\@.
  \end{minipage}
\end{table*}
Fig.~\ref{fig:etot}).  Here, $E_\text{int}$, $E_\text{kin}$ and
$E_\text{grav}$ are the global values of internal, kinetic and
gravitational binding energy, respectively.  The most energetic
explosion model (N150def) releases roughly seven times the energy of
the least energetic (N1def).  To understand the differences, the
temporal evolution of $E_\text{tot}$ and the energy release rate
$\dot{E}_\text{nuc}$ is shown in Fig.~\ref{fig:etot}.  In the upper
panel, all curves start from the initial binding energy $E_\text{bind}
= E_\text{tot}(t = 0) = -5.19 \times 10^{50}~\text{erg}$.  After
$1.7$~s, deflagration burning has ceased in all models and released
the energy $E_\text{def} = E_\text{nuc}(t =
1.7~\text{s})$.\footnote{\label{fn:additional_burning_lset}%
  In models with a bound remnant, some more burning occurs at later
  times during the pulsations of the remnant, as our level set based
  burning scheme is not turned off and still releases some energy if
  the (former) flame front is advected to or shocked to high
  densities.}  To illustrate how the different speeds of burning and
expansion lead to different explosion energies, Fig.~\ref{fig:massrho}
shows the distribution of burnt masses over the available fuel
densities.

\begin{figure}
  \centering
  \includegraphics[width=\columnwidth]{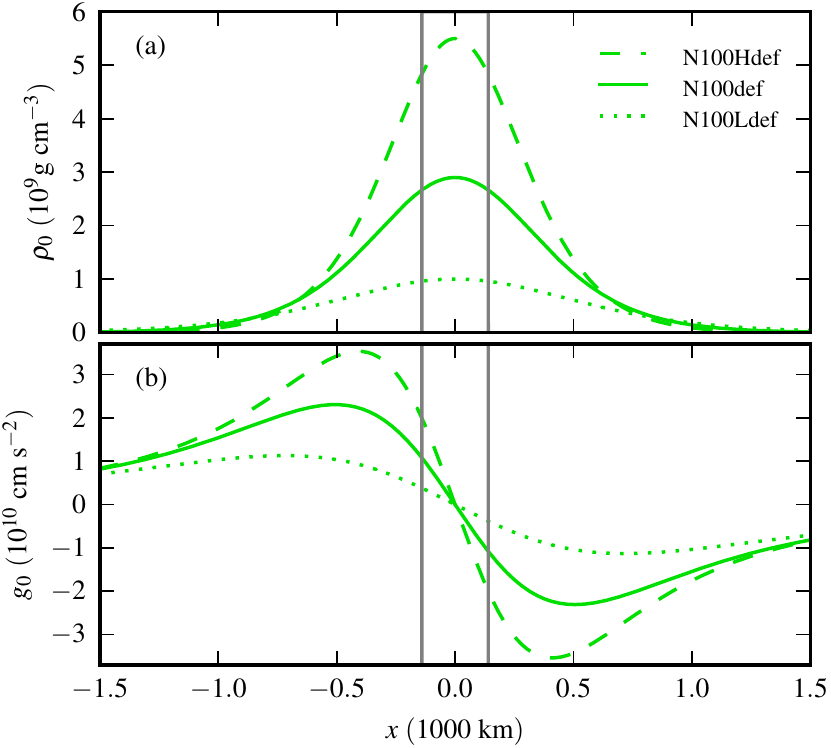}
  \caption{Initial density profiles (a) and gravitational acceleration
    (b) of the N100 deflagration models with different central
    densities.  The ignition kernels are located between the two
    vertical grey lines.}
  \label{fig:inipos_LH}
\end{figure}

Multispot ignition of SN~Ia models has been studied extensively in
previous work \citep[e.g.][]{roepke2006a}.  Here, we summarize the
main results to explain the energetics within our series of models.
We start with effects that lead to the acceleration of the burning.
An increase in the number of ignition kernels $N_\text{k}$ causes:
(i)~a growth of the initial flame surface area; (ii)~an increase of
the buoyancy force on the dominating outermost flame features (as
$r_\text{max}$ grows with $N_\text{k}$ for our standard
models);\footnote{The buoyancy force increases with the effective
  gravitational acceleration at the location of the bubble and with
  the bubble size (see, e.g., \citealt{roepke2006a}, for further
  details).  In the inner parts of our WDs, the absolute value of the
  gravitational acceleration increases with radius (see
  Fig.~\ref{fig:inipos_LH}b).}
and (iii)~the excitation of more modes of instabilities that produce
turbulence and increase the flame surface.  All these effects cause a
faster evolution/acceleration of the dominant flame features and thus
an increase in $\dot{E}_\text{nuc}$ with $N_\text{k}$ (see
Fig.~\ref{fig:etot}a).  As the burning competes with WD expansion,
an increase in $\dot{E}_\text{nuc}$ leads to a greater amount of burnt
mass at all densities (see Fig.~\ref{fig:massrho}a) and a higher
deflagration energy $E_\text{def}$.

In contrast to these accelerating effects, for large $N_\text{k}$,
$\dot{E}_\text{nuc}$ and $E_\text{def}$ decrease again due to
(iv)~early flame surface destruction (if neighbouring flame structures
are very close to each other and merge quickly) and (v)~rapid early WD
expansion.  Consequently, for low to intermediate $N_\text{k}$,
$E_\text{def}$ increases, reaches a maximum at $N_\text{k} = 150$ and
decreases again for $N_\text{k} > 150$.  Around $N_\text{k} = 150$,
the growth of the flame is well balanced with the rate of stellar
expansion: roughly equally sized flame structures are distributed over
the whole solid angle, which maximizes the amount of matter burnt
\citep[cf.][]{roepke2006a}.  Note that our optimum number of ignition
sparks differs significantly from the value found by
\citet{roepke2006a} (150 kernels \emph{per octant}), as they used
ignition kernels with a smaller radius $R_\text{k} = 5~\text{km}$
(this study uses $R_\text{k} = 10~\text{km}$ in all models except
N300Cdef).  Fig.~\ref{fig:massrho}(a) shows that at high densities
models with very high $N_\text{k}$ burn more than N150def, but at low
densities they burn significantly less.  Despite these differences, as
in \citet{roepke2006a}, all vigorously ignited models fall into a
narrow range of $E_\text{tot}$ around $6 \times 10^{50}~\text{erg}$ at
the end of the deflagration phase.

In models with \emph{compact} ignition configuration (dotted and
dashed lines in Fig.~\ref{fig:etot}), the same number of ignition
kernels is located within a smaller volume compared to the other
models in our sequence.  Model N1600Cdef has lower
$\dot{E}_\text{nuc}$ and $E_\text{def}$ than N1600def, due to reduced
buoyancy forces (ii) and even more severe flame surface destruction
(iv).
For model N300Cdef, both the significantly lower value of
$r_\text{max}$ and the lower ignition kernel radius, $R_\text{k} =
5~\text{km}$ (half the value of the other models) lead to very low
buoyancy forces (ii) and cause the burning in this model to evolve
significantly more slowly than in all others.  However, the smaller
value of $R_\text{k}$ allows the bubbles that are initially burning to
rise before they merge with their neighbours.  As a result, the
ignition configuration eventually evolves to a state similar to that
in the less compact configurations (e.g.\ N100def) and the total
energy release is comparable.

Recently, \citet{long2013a} have also studied 3D deflagration models
based on multispot ignition setups.  They find that their models with
fewer ignition kernels reach higher explosion energies than those with
more ignition kernels.  This relation holds only because most of their
setups with large $N_\text{k}$ are `saturated': they place as many
ignition kernels into the ignition volume as possible without
intersection.  Thus, effects of early flame surface destruction (iv)
and rapid early expansion (v) are very pronounced and lead to
relatively low values of $E_\text{def}$.  Among their models with
non-saturated setups, however, similar trends hold as for our set of
models ($E_\text{def}$ increases from $N_\text{k} = 63$ to 128 and
then decreases again for $N_\text{k} = 1700$).  Most of the setups in
our study are far from being saturated.

We have studied the importance of the central density for the N100
ignition geometry (see Fig.~\ref{fig:inipos_LH} for a comparison of
the three density profiles and the location of the ignition kernels).
The main difference of the `H' (`L') version is that the deflagration
starts off at significantly higher (lower) densities and that there is
more (less) high-density material available for burning.  This is
directly reflected in the density distribution of burnt matter (see
Fig.~\ref{fig:massrho}b): with respect to N100def, the right peak in
the H (L) model shifts to higher (lower) densities, whereas the
low-density side of the distribution does not show significant
changes.  Dynamically, the increased (decreased) mass that sits at $r
< r_\text{max}$ leads to an increase (decrease) of the gravitational
acceleration (see Fig.~\ref{fig:inipos_LH}b) and thus to more (less)
buoyancy.  Consistently, the flame evolves faster for higher central
densities (see Fig.~\ref{fig:etot}c).  Combined with the higher amount
of fuel at higher density, this also leads to higher deflagration
energies $E_\text{def}$.  While the L model has a significantly lower
final $E_\text{def}$ than the standard model, an increase of the
central density (as in the H model) does not yield a substantial
increase of $E_\text{def}$.

\subsection{Unbinding the white dwarf star}
\label{sec:unbinding_wd}

Deflagrations can leave a considerable mass of high-density unburnt
fuel close to the centre of the WD\@.  This material is not
accelerated to high speeds in the course of the explosion.  If, in the
end, the velocity of a volume element within this matter does not
reach the escape velocity with respect to the mass that sits at lower
radii, it stays bound and is not ejected.  To determine which part of
the WD becomes unbound, we have calculated the asymptotic specific
kinetic energy, $\epsilon_\text{kin,a}$, as the sum of the specific
gravitational and kinetic energies for each cell on our hydrodynamic
grid at the end of the simulation:\footnote{We have neglected any
  potential contribution of the internal energy here.  Its inclusion
  does not lead to a significant decrease of the bound masses (at most
  $1.8$ per cent) of our model sample.}
\begin{equation}
  \label{eq:etot}
  \epsilon_\text{kin,a} = \epsilon_\text{grav}(100~\text{s}) +
  \epsilon_\text{kin}(100~\text{s}).
\end{equation}
For $\epsilon_\text{kin,a} > 0$, a cell becomes unbound and will
eventually approach the asymptotic velocity
\begin{equation}
  \label{eq:vel_a}
  v_\text{a} = \sqrt{2 \epsilon_\text{kin,a}}.
\end{equation}
If $\epsilon_\text{kin,a} \le 0$, the cell will be left behind after
the explosion and stay in a bound remnant.  Note that models with
similar total energy release can have significantly different ejecta
masses since it is crucial where the explosion energy is deposited.

\begin{figure}
  \centering
  \includegraphics[width=\columnwidth]{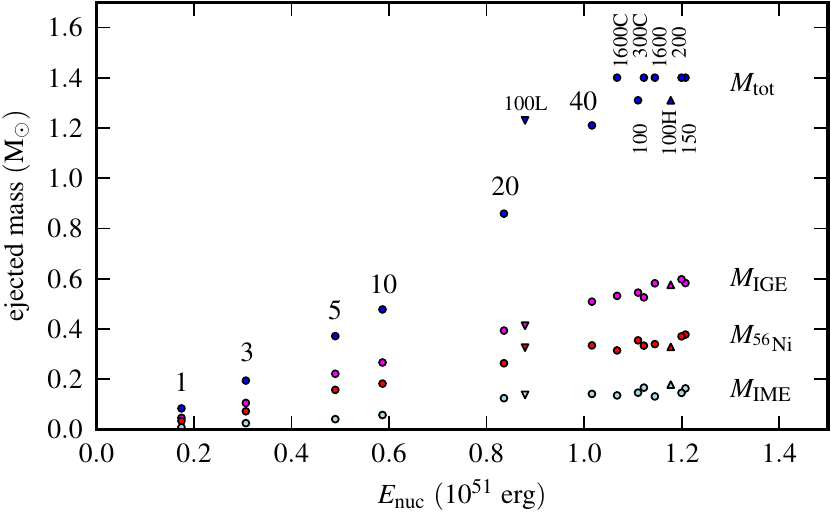}
  \caption{Total ejected masses and IGE, $^{56}$Ni and IME masses
    among the ejecta versus nuclear energy release in the explosion.
    Models with ${\leq} 100$ ignition sparks leave a compact remnant.}
  \label{fig:unbound_mass}
\end{figure}

We have determined ejecta masses $M_\text{ej}$ and masses of the bound
remnants $M_\text{b}$ for all of our models (see
Table~\ref{tab:energetics} and Fig.~\ref{fig:unbound_mass}).
Models with many ignition kernels ($N_\text{k} > 100$) release a large
amount of energy and deposit sufficient energy close to the centre
(e.g.\ by transferring kinetic energy to the downdrafts of unburnt
fuel) to unbind the whole WD, including the central unburnt fuel.  In
the case of a low or intermediate number of ignition sparks
($N_\text{k} \lesssim 100$), only part of the WD becomes unbound.
Most of the unburnt fuel remains in the remnant, while the ejected
part reaches homologous expansion.  We find a continuum of remnant
masses between $1.32$ and $0.09~\msol$ and, as
Fig.~\ref{fig:unbound_mass} shows, a monotonic (roughly linear)
increase of $M_\text{ej}$ with the nuclear energy release
$E_\text{nuc}$ below $E_\text{nuc} \sim 1.1 \times 10^{51}$~erg (or
for $N_\text{k} < 100$).  Table~\ref{tab:energetics} also gives the
values of the asymptotic kinetic energies of the ejecta.  Further
properties of the remnants are discussed in Section~\ref{sec:remnant}.

\subsection{Pulsations/chances of a secondary detonation?}
\label{sec:puls_secdet}

\begin{figure}
  \centering
  \includegraphics[width=\columnwidth]{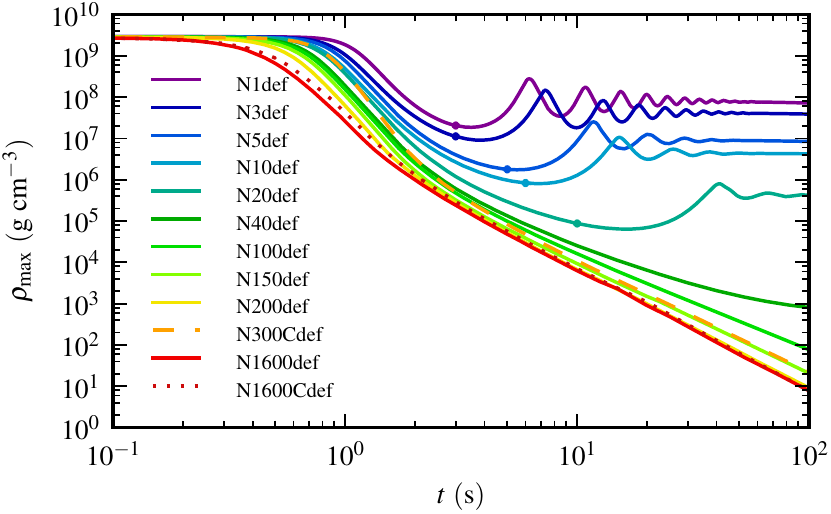}
  \caption{Maximum density $\rho_\text{max}$ on the hydrodynamic grid
    as indicator for the temporal evolution of expansion and
    pulsations.  For the five weakest deflagration models, the
    evolution of $\rho_\text{max}$ at late times was determined in
    separate simulations in which the grid expansion was stopped to
    keep the bound remnants spatially resolved (parts of the curves
    after the circle marks).}
  \label{fig:pulsations}
\end{figure}

As reported by other recent studies \citep{bravo2009a, jordan2012b},
the weak deflagrations in our model series also lead to pulsations
within the WDs: Fig.~\ref{fig:pulsations}
shows that the burning causes significant expansion (seen in the
figure as a decrease in the temporal evolution of the maximum
density).  For models with low $N_\text{k}$, the bound inner parts
start to contract after the deflagration has ceased.  After maximum
compression, the dense core region starts to expand again, while outer
parts are still falling inwards.  Thus, an accretion shock forms
somewhere near the edge of the dense core (for a more detailed
description, see e.g. \citealt{bravo2009a}) and some matter around
this shock region is heated up significantly.  Shortly thereafter, the
infall is stopped and all matter is again moving outwards, but, as it
is still gravitationally bound, further pulsations ensue with steadily
decreasing amplitudes.  The weaker the strength of the deflagration
and the more massive the bound remnant of a model, the more violent
are these pulsations.

Previous studies found that the high temperatures in the accretion
shock may be sufficient to trigger a secondary detonation for
$E_\text{nuc} \lesssim E_\text{bind}$ \citep{bravo2009a}.  Since the
ignition occurs at the radius of the accretion shock,
\citeauthor{bravo2009a} call this scenario pulsating reverse
detonation (PRD).  If the deflagration is very one sided, however,
even before the formation of an accretion shock, a hotspot may form
due to the deflagration ashes converging on the opposite side of the
ignition.  This so-called classical GCD scenario typically works for
low deflagration energies with $E_\text{nuc} \sim 0.1 E_\text{bind}$
\citep{jordan2012a}.  For slightly stronger deflagrations (which cause
significant stellar expansion in the early burning stages), the
conditions may be insufficient for detonation initiation at this
point.  But, as \citet{jordan2012a} report, for $E_\text{nuc}$ within
$0.38$--$0.78 E_\text{bind}$, the compression at the onset of the
first pulsation may be sufficient to increase both temperature and
density in the hotspot to the critical values (see
\citealt*{roepke2007a} and \citealt{seitenzahl2009c}).  This so-called
pulsationally assisted GCD will likely precede a potential PRD\@.

In our series of models, we do not find conditions suitable for any
secondary detonation for models with $N_\text{k} \ge 5$.  Model N5def
just reaches $10^9$~K in the accretion shock, but only at densities
which are too low for detonation initiation ($\rho < 10^6~\gcc$).  All
stronger deflagrations ($N_\text{k} \ge 10$) do not even reach
temperatures of $10^9$~K, which would be necessary for detonation.
For our weakest deflagration models N1def and N3def, we cannot exclude
a secondary detonation.  Their explosion energies of $0.34$ and $0.59
E_\text{bind}$
could make them candidates for potential pulsationally assisted GCDs
or PRDs.  However, as we find conditions that are only marginally
critical (according to \citealt{roepke2007a} and
\citealt{seitenzahl2009b}), only future studies with simulations
designed to sufficiently spatially resolve the relevant regions can
provide an answer.  Here, we focus on the outcomes of pure
deflagration models.

\subsection{Properties of the remnant object}
\label{sec:remnant}

As discussed in Section~\ref{sec:unbinding_wd}, our weakest
deflagrations leave behind bound remnants with masses given in
Table~\ref{tab:energetics}.  Both the remnant mass and the strength of
the pulsations increase with decreasing nuclear energy release of the
deflagration.  After the first and strongest pulsation, a dense nearly
hydrostatic \citep[cf.][]{bravo2009a} core forms within the bound
material.  This core, which is heated in the pulsations and the
accretion shock, is similar to a very hot WD and contains most of the
bound mass.  The outer layers of the core are already enriched with
deflagration ashes as the innermost burnt regions have never become
unbound.  The bound mass outside the core is also a mix of
deflagration ashes and unburnt fuel; it may eventually settle down on
the hydrostatic core.  For a detailed description of the
nucleosynthetic yields in the bound remnants, see
Section~\ref{sec:total_nucl_core}.

In asymmetric explosions that leave remnants, a recoil momentum of the
remnant may be expected.  \citet{jordan2012b} report a high kick
velocity of a few hundred \kms\ for their deflagration models that
were ignited at off-centre points.  Such strong kicks could be
sufficient to eject the remnants from the system and produce
potentially observable runaway/hypervelocity WDs.  However, in our
study we do not find such high kick velocities (see
Table~\ref{tab:kick_vel}):
\begin{table}
  \centering
  \begin{minipage}{67.5mm}
    \caption{Centre-of-mass kick velocities of the bound remnants.}
    \label{tab:kick_vel}
    \begin{tabular}{@{}lclc@{}}
      \hline
      Model & $v_\text{kick}~(\kms)$ & Model & $v_\text{kick}~(\kms)$ \\
      \hline
      N1def & 5.1 & N10def & 6.8 \\
      N1def\_FFT & 8.2 & N20def & 18 \\
      N3def & 4.4 & N40def & 32 \\
      N5def & 5.4 & N100def & 16 \\
      N5def\_FFT & 36 \\
      \hline
    \end{tabular}
  \end{minipage}
\end{table}
our values are at maximum 5 to 10 per cent of the values in
\citet{jordan2012b}.  This might be partially due to a higher degree
of asymmetry in the ignition setups of \citeauthor{jordan2012b}
(compared to our sparsely ignited models), who ignite (in most cases)
relatively large off-centre volumes of radius 128~km.  On the other
hand, our monopole gravity solver, which is used in most models, could
suppress higher kick velocities.  To test the role of the gravity
solver, we have re-simulated our models N1def and N5def using a fast
Fourier transformation-based gravity solver that solves the full 3D
Poisson equation without approximations (see Section~\ref{sec:hd}).
Currently, this solver is restricted to uniform grid geometries.
Thus, these models have a lower spatial resolution of the inner parts
of the WD than our other models (which use a hybrid grid; see
Section~\ref{sec:hd}).  Due to the lower spatial resolution of the
initial flame, these new models N1def\_FFT and N5def\_FFT release
somewhat less energy in the explosion, but qualitatively they agree
well with the corresponding monopole gravity simulations and the
large-scale flame evolution and asymmetry are very similar.  Regarding
the kick velocity, only N5def\_FFT has a noticeably higher value
(36~\kms), but this is still roughly of an order of magnitude lower
than the values found by \citet{jordan2012b}.  The main results of
this work, namely the nucleosynthesis and the observable predictions
for the ejected matter should, however, not depend on a potential
recoil.

\section{Nucleosynthesis}
\label{sec:nucl}

In this section, we discuss the results of our detailed
nucleosynthesis post-processing calculations (see Section~\ref{sec:pp}).
We first cover the total integrated yields within the ejecta and the
remnant objects (Sections~\ref{sec:total_nucl} and
\ref{sec:total_nucl_core}) and then present the detailed geometrical
structures of the ejected nucleosynthetic yields in velocity space
(Section~\ref{sec:vel_nucl}).

\subsection{Total nucleosynthetic yields in the ejecta}
\label{sec:total_nucl}

Global yields of stable and radioactive nuclei in the ejecta are shown
in Tables~\ref{tab:syields} and \ref{tab:ryields} in the Appendix.
For a quick overview, yields of the most important (classes of)
species are also given in Table~\ref{tab:energetics} and plotted in
Fig.~\ref{fig:unbound_mass}.

As shown in Fig.~\ref{fig:unbound_mass} and explained in
Section~\ref{sec:expl_ene}, for $N_\text{k} \le 150$ the nuclear energy
release $E_\text{nuc}$ increases with the number of ignition bubbles
$N_\text{k}$; for larger $N_\text{k}$, $E_\text{nuc}$ decreases.  The
mass of $^{56}$Ni produced scales with $E_\text{nuc}$ and also follows
these trends.  The maximum $^{56}$Ni mass of $0.38~\msol$, which is
produced in model N150def, is marginally compatible with the lower end
of the distribution of $^{56}$Ni masses reported for normal SNe~Ia
\citep[cf.\ e.g.][]{stritzinger2006a}.  Neutronization by electron
captures is only efficient at the highest burning densities.
Consequently, the degree of neutronization increases with $N_\text{k}$
in our model series since, as shown in Fig.~\ref{fig:massrho}(a),
models with higher $N_\text{k}$ burn more mass at the highest fuel
densities.  The ratio of the mass of $M_{^{56}\text{Ni}}$ (which has
equal numbers of protons and neutrons) to the total mass of IGEs that
is produced in NSE thus also goes down with $N_\text{k}$ (see
Table~\ref{tab:energetics}).  Models N100Ldef and N100Hdef are
outliers from this general trend: due to their lower/higher central
densities, they have a lower/higher degree of neutronization and thus
$^{56}$Ni masses that lie slightly above/below the mean trend in
Fig.~\ref{fig:unbound_mass} \citep*[see also][]{seitenzahl2011a}.  The
total mass of IMEs (e.g.\ $^{28}$Si, $^{32}$S) is roughly $20$ to $30$
per cent of the total mass of IGEs.  One of the less obvious results
is the decrease of the amount of burning products (estimated roughly
by $\frac{M_\text{IGE} + M_\text{IME}}{M_\text{ej}}$ in
Table~\ref{tab:energetics}) within the ejecta from weak to strong
deflagrations: the mass fraction of burning products ranges from about
$70$ down to only $50$ per cent.  This occurs because, for weak
deflagrations, a large fraction of the unburnt fuel becomes part of
the bound remnant and is thus not released into the ejecta.

\subsection{Nucleosynthetic yields in the bound remnant}
\label{sec:total_nucl_core}

Total integrated nucleosynthetic yields of stable and radioactive
nuclei in the bound remnants are given in
Tables~\ref{tab:syields_core} and \ref{tab:ryields_core} in the
Appendix.  An overview of the composition is provided in
Table~\ref{tab:core_comp}:
\begin{table*}
  \centering
  \begin{minipage}{129mm}
    \caption{Composition of the bound remnants (masses in solar
      masses).}
    \label{tab:core_comp}
    \begin{tabular}{@{}lccccccccc@{}}
      \hline
      Model & $M_\text{b}$ & $M_{^{56}\text{Ni}}$ & $M_\text{IGE}$ & $M_\text{IME}$ &
      $M_{^{20}\text{Ne}}$ & $M_{^{16}\text{O}}$ & $M_{^{12}\text{C}}$ &
      $\frac{M_{^{56}\text{Ni}}}{M_\text{IGE}}$ & $\frac{M_\text{IGE} + M_\text{IME}}{M_\text{b}}$ \\
      \hline
      N1def    &  1.32 &  0.0325 &  0.0417 &  0.0354 &  0.0448 &  0.622 &  0.547 & 0.78 & 0.058 \\
      N3def    &  1.21 &  0.0159 &  0.0230 &  0.0391 &  0.0461 &  0.578 &  0.497 & 0.69 & 0.051 \\
      N5def    &  1.03 &  0.0221 &  0.0305 &  0.0352 &  0.0345 &  0.486 &  0.424 & 0.72 & 0.064 \\
      N10def   & 0.926 &  0.0214 &  0.0296 &  0.0413 &  0.0360 &  0.432 &  0.369 & 0.72 & 0.077 \\
      N20def   & 0.545 & 0.004\,45 & 0.006\,21 &  0.0175 &  0.0235 &  0.261 &  0.225 & 0.72 & 0.044 \\
      N40def   & 0.190 & 0.000\,80 & 0.001\,26 & 0.004\,04 & 0.008\,72 & 0.0921 & 0.0799 & 0.63 & 0.028 \\
      N100Ldef & 0.133 & 0.002\,62 & 0.003\,18 & 0.003\,49 & 0.000\,74 & 0.0637 & 0.0584 & 0.82 & 0.050 \\
      N100def  & 0.090 & 0.000\,76 & 0.001\,05 & 0.001\,99 & 0.002\,02 & 0.0437 & 0.0392 & 0.73 & 0.034 \\
      N100Hdef & 0.102 & 0.000\,38 & 0.000\,61 & 0.009\,55 &  0.0181 & 0.0462 & 0.0258 & 0.63 &  0.10 \\
      \hline
    \end{tabular}
  \end{minipage}
\end{table*}
as previously reported by \citet{jordan2012b} and discussed in
Section~\ref{sec:remnant}, the bound remnants produced in the weak
deflagration models are enriched by products of the explosive burning
(mainly in the outer layers).  If we count only IGEs and IMEs, these
products contribute 3--10 per cent to the remnant mass (see
Table~\ref{tab:core_comp}).  After cooling down, the remnant objects
will again become WDs, but with very peculiar composition.

In addition to the deflagration products, some oxygen and neon might
also have been produced by carbon burning during the pulsations of the
central regions (densities are too low to reach NSE).  In this study,
we have not incorporated the energy release of this burning phase
correctly in the hydrodynamic simulations and so may somewhat
underestimate the burning yields in our post-processing (see, however,
footnote~\ref{fn:additional_burning_lset}).

As discussed in detail by \citet{kromer2013a}, radioactive material in
the outer layers of the bound remnants may explain the relatively
slowly declining light curves and peculiarities in the late-time
spectra of 2002cx-like SNe \citep{jha2006b, phillips2007a}.
Currently, radioactive decays in the bound remnants are not taken into
account by our radiative transfer simulations.

\subsection{Velocity distribution of ejected yields}
\label{sec:vel_nucl}

The distribution of nucleosynthetic yields in velocity space and the
underlying density profile determine the observable outcomes of the
explosion.  We have calculated synthetic light curves and spectra for
our models (see Section~\ref{sec:synth_obs}).  For a better
understanding of the results, we describe the most important
properties of the nucleosynthetic yield distributions here.

\begin{figure*}
  \centering
  \includegraphics[width=\textwidth]{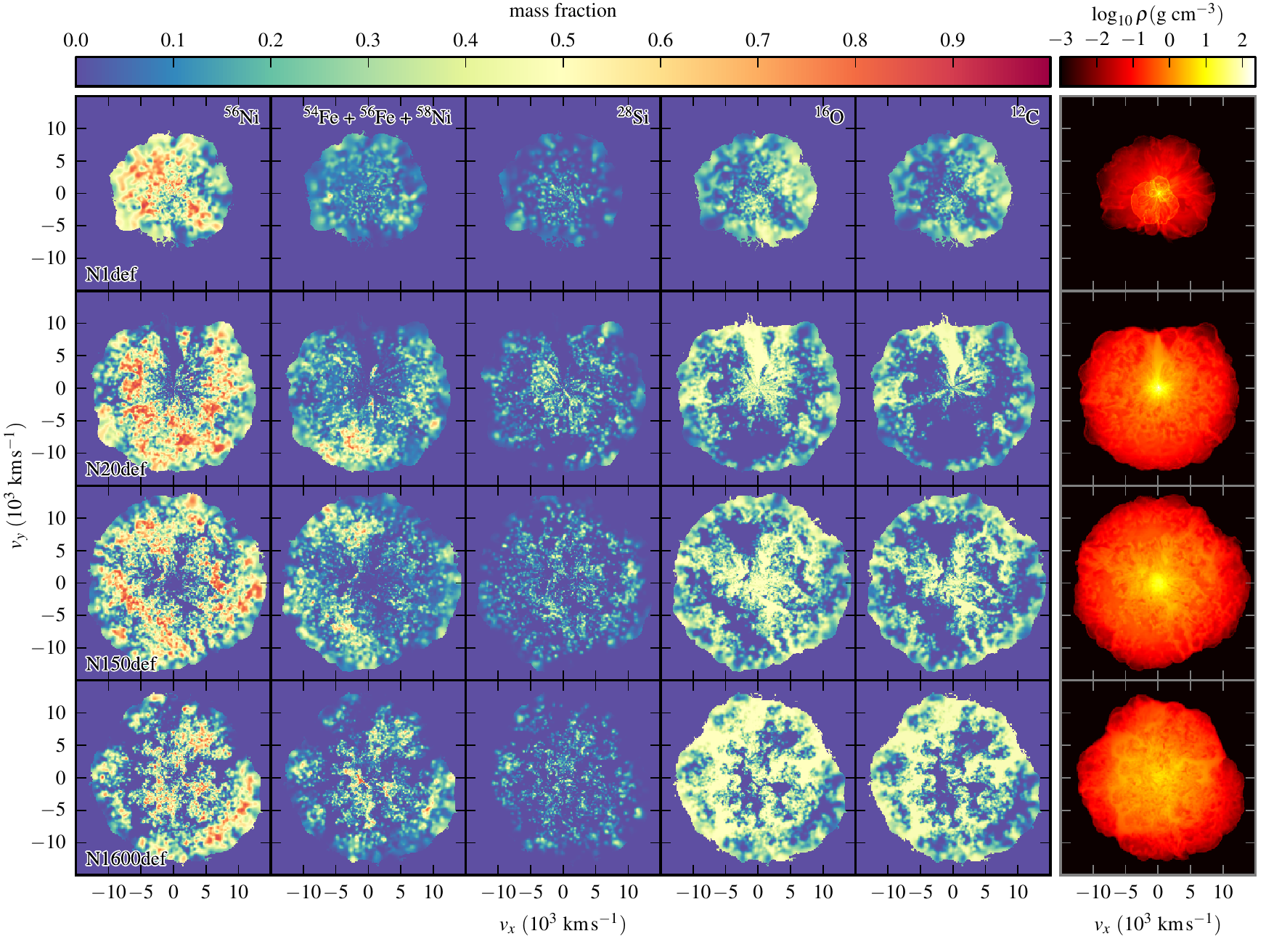}
  \caption{Slices through the mid-plane of the final ejecta
    distributions of the models N1def, N20def, N150def and N1600def
    (top to bottom) in asymptotic velocity space (i.e.\ only material
    reaching escape velocity is shown).  Colour coded are the mass
    fractions of $^{56}$Ni, the sum of the most abundant stable IGEs
    ($^{54}$Fe, $^{56}$Fe and $^{58}$Ni), the IME $^{28}$Si, the
    remaining $^{16}$O and $^{12}$C fuel and $\log_{10} \rho$ (left to
    right).  The abundance distributions have been calculated by
    mapping the tracer particle distributions on to a $200^3$
    Cartesian grid as described in Section~\ref{sec:rt}.  For the
    density, the cells of the hydrodynamic grid that reach escape
    velocity have been mapped in the same way as the tracer
    particles.}
  \label{fig:map_abund}
\end{figure*}

\begin{figure}
  \centering
  \includegraphics[width=\columnwidth]{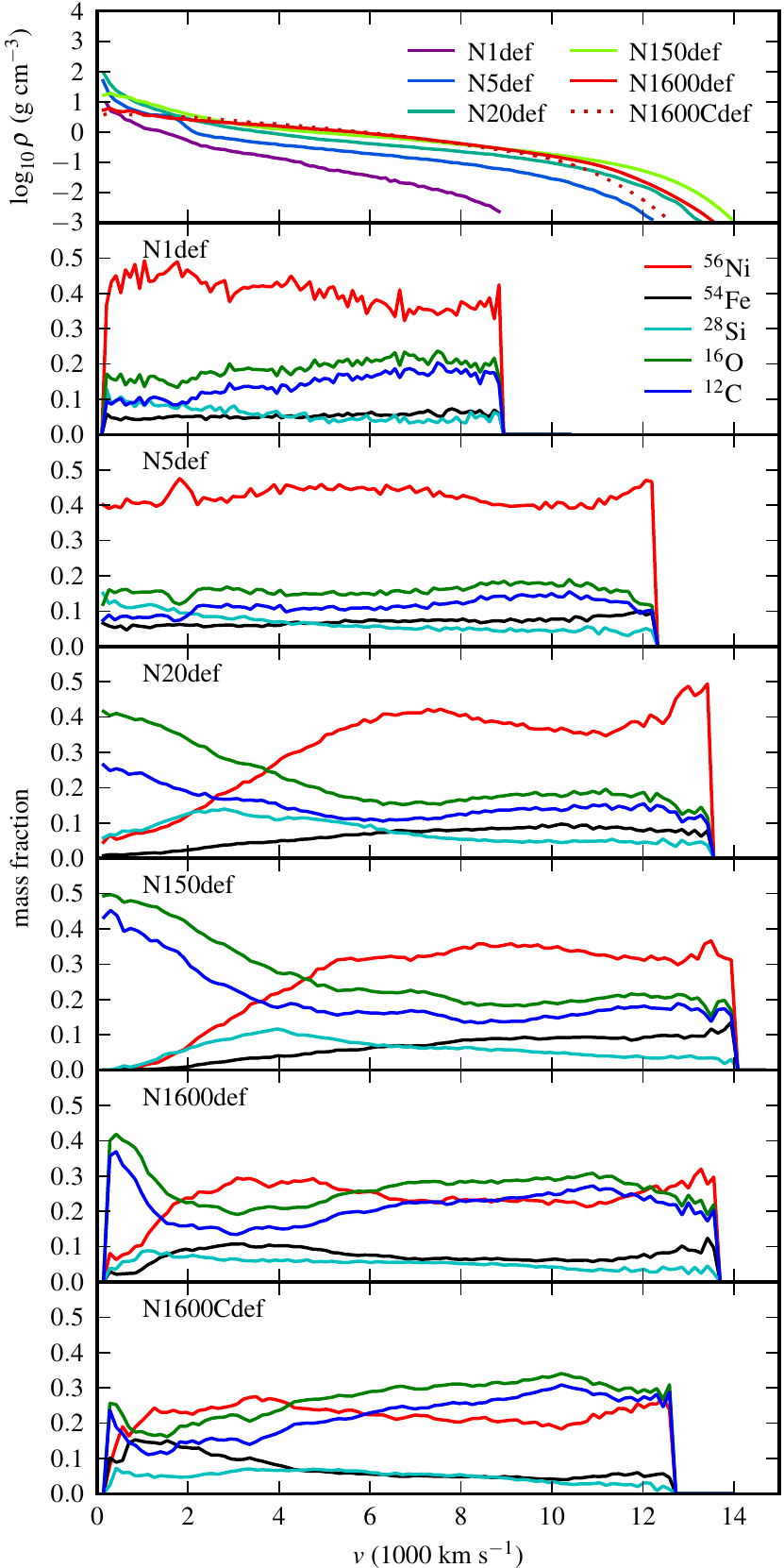}
  \caption{Angle-averaged density and composition in asymptotic
    velocity space (i.e.\ only material reaching escape velocity is
    shown).  For the averaging, 100 radial bins were used.}
  \label{fig:tomo1}
\end{figure}

Fig.~\ref{fig:map_abund}
shows the final abundance distributions of select species and the
density in 2D slices in asymptotic velocity space.  Again, we focus
our discussion on the four models N1def, N20def, N150def and N1600def,
which are representative for the whole sample and its variety (see
Fig.~\ref{fig:tomo1}
for spherically averaged distributions).  The distributions show
complex multidimensional structures typical for deflagration models.
The outer contours of the ashes of the main burning plumes are most
clearly visible in the $^{12}$C abundance plots (as regions where
carbon is absent; cf.\ also the final snapshots in
Figs~\ref{fig:expl_evol} and \ref{fig:expl_evol2}).  Inside the
ash-rich regions, the typical products of burning in C/O matter are
found: at the outer edges, incomplete burning leads to the production
of $^{16}$O and IMEs including $^{28}$Si (which is nowhere very
prominent).  The main burning products are, however, IGEs.  The stable
IGEs, which are mostly neutron rich, are produced in the initial
stages of the deflagration, when the burning bubbles are still located
at the high-density regions close to the centre.  Heated by the
burning, the ashes (and also surrounding unburnt fuel) rise up through
buoyancy.  In this way, the burning at the edges of the deflagration
plumes burns the matter only after some pre-expansion.  At these lower
densities, the freeze-out composition from NSE is mainly $^{56}$Ni.
Therefore, the $^{56}$Ni in the distribution plots mostly surrounds
the regions rich in stable IGEs.  Due to the inhomogeneous growth and
structure of the deflagration plumes, there are also inhomogeneities
in the distribution of IGEs.  Only if all IGEs are summed do the large
contiguous red areas seen in Figs~\ref{fig:expl_evol} and
\ref{fig:expl_evol2} become visible.

Between the burnt structures, the ejecta are filled with unburnt
material that has sunk down between the burning bubbles or has
expanded less than the hot ashes.  As discussed above, one-sided
deflagrations (typical for low $N_\text{k}$) tend to extinguish before
the burning front can completely wrap around the outer layers of the
whole star.  Therefore, a channel of unburnt matter that reaches from
the ejecta surface down to the centre is present in several models
(e.g.\ N20def).  Models with intermediate numbers of ignition sparks
($N_\text{k} \sim 40$--$150$) tend to have relatively few large
equally sized contiguous burning plumes.  Unburnt fuel is found mainly
close to the centre and in thin channels between the plumes.  Models
with larger numbers of ignition sparks ($N_\text{k} \ge 200$) tend to
have a greater number of smaller burning plumes.  Regions of unburnt
fuel occupy a very large volume and are most prominent in the outer
parts of the ejecta (thus models like N1600def show large mass
fractions of unburnt fuel at high velocities in Fig.~\ref{fig:tomo1}).
For all models, the unburnt structures can also be seen as regions
with slightly higher density than neighbouring burnt material, as they
have expanded less (rightmost panels of Fig.~\ref{fig:map_abund}).  In
model N1def (and the other models with $N \le 10$) the former
accretion shock of the most violent pulsation is still visible as
asymmetric shell-like structure with a jump in density.

Despite the small-scale asymmetries due to turbulent mixing, the
large-scale asymmetries of the final ejecta structures are only
moderate.  Often, asymmetries during the early burning stages are
mitigated later, when the burning plumes expand into unburnt regions.
As explained above, sparsely ignited one-sided deflagrations tend to
be especially asymmetric and models with intermediate $N_\text{k}$ are
more symmetric than all other models.  Observable consequences of such
large-scale asymmetries will be discussed in
Section~\ref{sec:synth_obs_los}.

\begin{figure}
  \centering
  \includegraphics[width=\columnwidth]{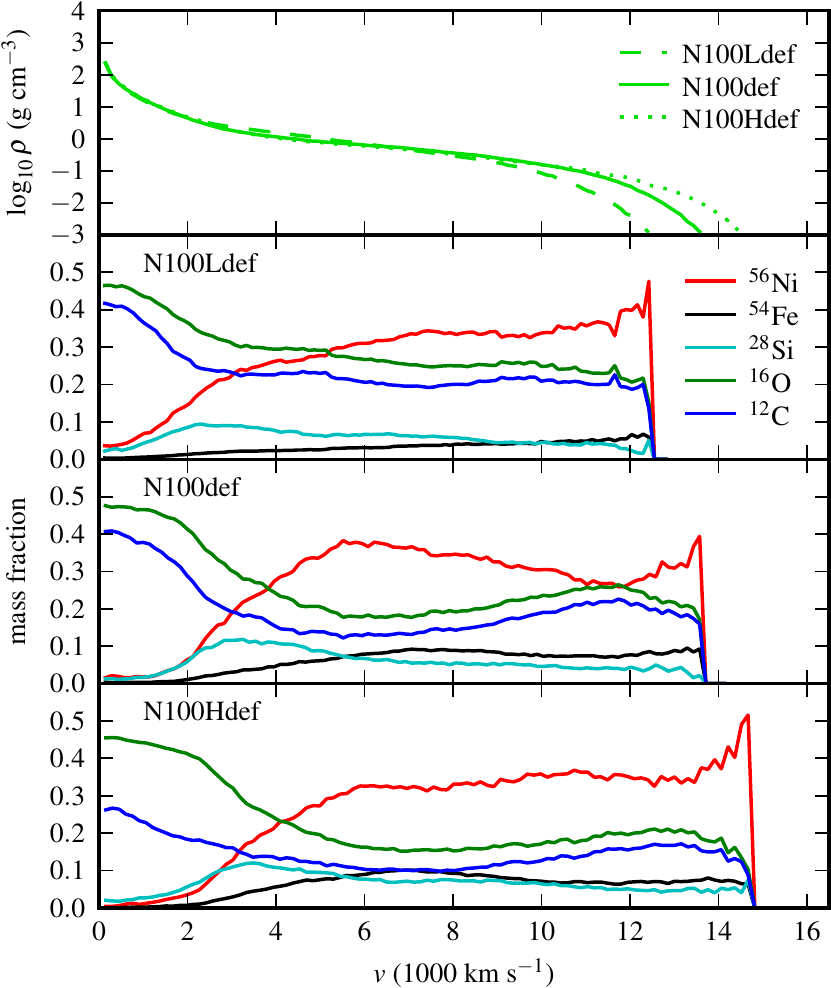}
  \caption{Same as Fig.~\ref{fig:tomo1}, but for the N100 deflagration
    models with different central densities.}
  \label{fig:tomo3}
\end{figure}

In the complex 3D structures of our deflagration models, burning
products such as $^{56}$Ni, stable IGEs and IMEs and also unburnt
$^{12}$C and $^{16}$O can be found at all ejecta velocities.
Spherical averages of our models show low maximum ejecta velocities
$v_\text{max}$ in the range $9000$--$15\,000~\text{km} \,
\text{s}^{-1}$ (see Figs~\ref{fig:tomo1} and \ref{fig:tomo3}).
The distribution of $v_\text{max}$ simply follows that of the (low)
explosion energies (see Section~\ref{sec:expl_ene} and
Fig.~\ref{fig:unbound_mass}).  We caution that spherical averages
suggest that the ejecta are very homogeneously mixed, which is, as we
have just seen, not really the case in multidimensional space.

In the weakest deflagration models, most of the unburnt material is
not ejected.  This unburnt material becomes part of the bound remnant.
Therefore, these models have a particularly homogeneous abundance
distribution in the ejecta (see models N1def and N5def in
Fig.~\ref{fig:tomo1}).

In models for which the ejected mass exceeds the bound mass
($N_\text{k} \ge 20$), unburnt material becomes increasingly dominant
in the central ejecta.  $^{56}$Ni and other IGEs are relatively scarce
in these central ejecta.  The adjacent layer with an almost
homogeneous mix of fuel and deflagration ashes (as seen in N1def and
N5def) is shifted towards higher velocities, compared to the weaker
models.

For models with increasing deflagration strength (and $N_\text{k}$ up
to $150$), the central region that is dominated by unburnt C/O
material extends to higher velocities.  For models with $N_\text{k}
\ge 200$, however, the amount of unburnt matter in the centre
decreases, as more and more high-density material is burnt in the
earliest burning stages (see Section~\ref{sec:expl_ene},
Fig.~\ref{fig:massrho}a).  Accordingly, the fraction of stable IGEs is
increased and extends down to the lowest velocities.  This behaviour
might be a side effect of our parametrization of the deflagration
strength with the multispot ignition scenario.

Partially bound models show a characteristic peak at the centre in
their density profile, whereas fully unbound models with $N_\text{k}
\gtrsim 200$ are flat.  This is consistent with the enhanced nuclear
burning and energy deposition of these models at the highest
densities.

The variation of the central density in models with the N100 ignition
geometry has only a moderate influence on the final ejecta structure
(Fig.~\ref{fig:tomo3}).  The qualitative changes for increasing
$\rho_\text{c}$ are similar to the changes for increasing $N_\text{k}$
along our series (for intermediate $N_\text{k}$).

\section{Synthetic observables}
\label{sec:synth_obs}

In this section, we present synthetic light curves and spectra for our
models as obtained from radiative transfer calculations with the
\textsc{artis} code (for a description of the simulation setup see
Section~\ref{sec:rt}).

\subsection{Light curves}
\label{sec:synth_obs_lcs}

$B$-band rise times ($t_\text{max}^B$), decline rate parameters
($\Delta m_{15}^B$) and peak absolute magnitudes of our angle-averaged
light curves for the full model sample are given in
Table~{\ref{tab:lc_props}}.
\begin{table*}
  \centering
  \begin{minipage}{96.5mm}
    \caption{Angle-averaged light curve time-scales (in days) and peak optical
      magnitudes for our models.}
    \label{tab:lc_props}
    \begin{tabular}{@{}lccccccc@{}}
      \hline
      Model & $t_\text{max}^B$ & $\Delta m_{15}^B$ & $M_\text{max}^U$ & $M_\text{max}^B$ & $M_\text{max}^V$ & $M_\text{max}^R$ & $M_\text{max}^I$ \\
      \hline
      N1def  & $7.6$          & $2.15$            & $-16.96$      & $-16.55$      & $-16.84$      & $-16.73$      & $-16.80$   \\
      N3def  & $9.6$          & $1.91$            & $-17.55$      & $-17.16$      & $-17.52$      & $-17.48$      & $-17.53$   \\
      N5def  & $11.1$         & $1.69$            & $-18.25$      & $-17.85$      & $-18.24$      & $-18.16$      & $-18.17$   \\
      N10def  & $11.1$        & $1.68$            & $-18.33$      & $-17.95$      & $-18.38$      & $-18.36$      & $-18.42$   \\
      N20def  & $12.0$        & $1.56$            & $-18.64$      & $-18.24$      & $-18.68$      & $-18.73$      & $-18.84$   \\
      N40def  & $12.9$        & $1.30$            & $-18.70$      & $-18.34$      & $-18.86$      & $-18.96$      & $-19.08$   \\
      N100Ldef& $13.8$        & $1.22$            & $-18.80$      & $-18.39$      & $-18.77$      & $-18.80$      & $-18.92$   \\
      N100def & $13.1$        & $1.31$            & $-18.75$      & $-18.40$      & $-18.92$      & $-19.03$      & $-19.16$   \\
      N100Hdef& $11.5$        & $1.44$            & $-18.69$      & $-18.34$      & $-18.89$      & $-18.99$      & $-19.12$   \\
      N150def  & $12.5$       & $1.28$            & $-18.81$      & $-18.43$      & $-18.96$      & $-19.10$      & $-19.25$   \\
      N200def  & $13.8$       & $1.04$            & $-18.59$      & $-18.26$      & $-18.89$      & $-19.07$      & $-19.22$   \\
      N300Cdef  & $12.0$      & $1.24$            & $-18.64$      & $-18.25$      & $-18.79$      & $-18.97$      & $-19.14$   \\
      N1600def  & $14.0$      & $0.95$            & $-18.44$      & $-18.11$      & $-18.76$      & $-18.97$      & $-19.13$   \\
      N1600Cdef & $14.4$      & $0.94$            & $-18.36$      & $-18.02$      & $-18.62$      & $-18.86$      & $-19.03$   \\
      \hline
    \end{tabular}
  \end{minipage}
\end{table*}

\begin{figure*}
  \centering
  \includegraphics[width=\textwidth]{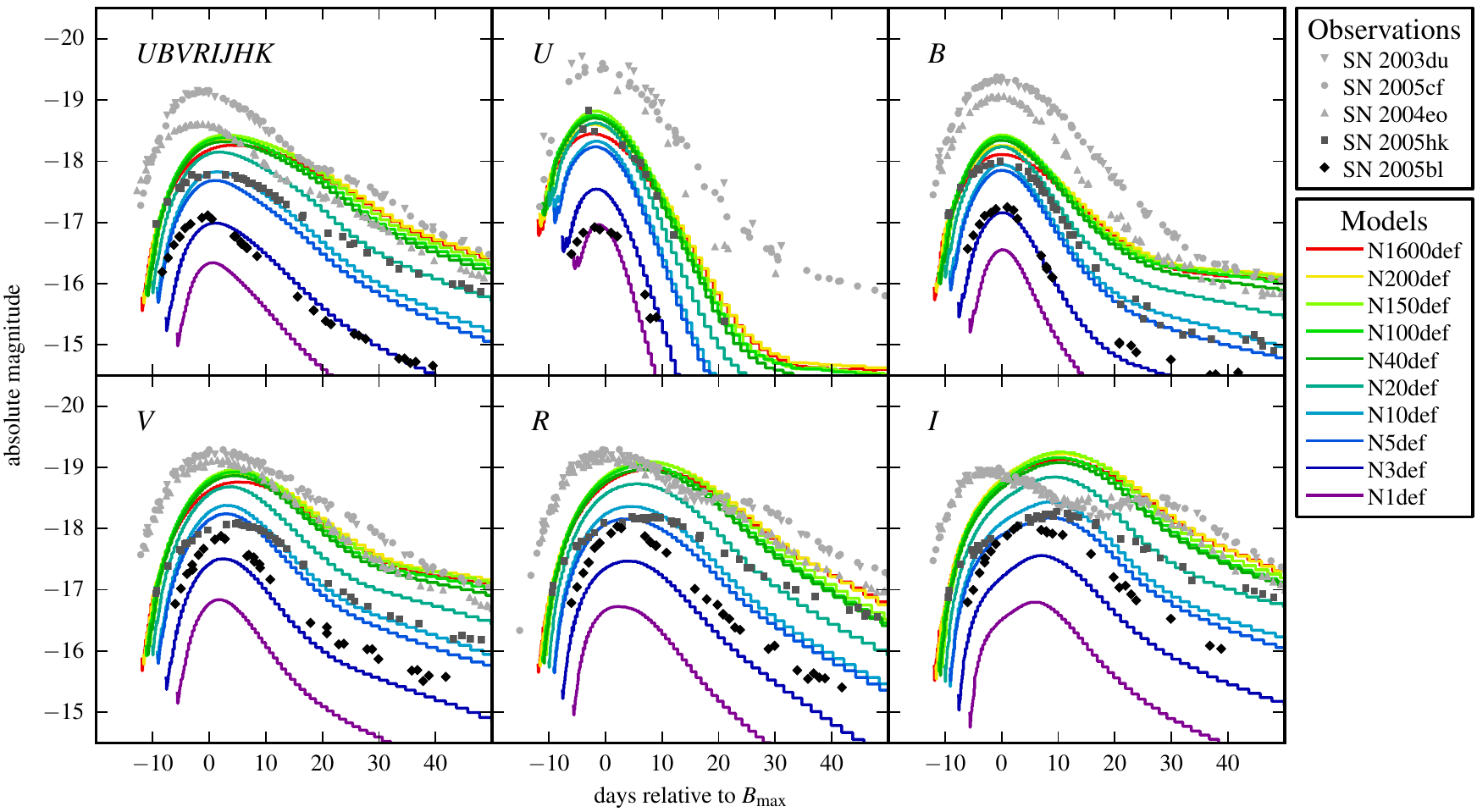}
  \caption{Angle-averaged synthetic light curves for select models of
    our sample (colour coded).  For comparison, light curves of
    several well-observed SNe~Ia are overplotted as black/grey
    symbols: SN~2003du, 2004eo and 2005cf
    \citep[][respectively]{stanishev2007b, pastorello2007b,
      pastorello2007a} representing normal SNe~Ia, SN~2005hk
    \citep{phillips2007a} representing 2002cx-like SNe and SN~2005bl
    \citep{taubenberger2008a} as a proxy for 1991bg-like SNe.}
  \label{fig:lc}
\end{figure*}

Angle-averaged bolometric and broad-band light curves for select
standard models are shown in Fig.~\ref{fig:lc}
and compared to a set of observed SNe~Ia.  As discussed in detail by
\citet{kromer2013a}, model N5def, which leaves behind a bound remnant
of $1.03~\msol$, reproduces the observed light curves of SN~2005hk
\citep{phillips2007a}, a prototypical event of the class of faint SNe
similar to SN~2002cx \citep{li2003a, jha2006b}, remarkably well.
Model N10def, which leaves behind a bound remnant of $0.926~\msol$,
yields light curves fairly similar to N5def, indicating a restricted
range of ignition conditions similar to those of N5def and N10def for
objects like SN~2005hk or SN~2002cx.

In contrast, models with a more vigorous ignition than N10def (i.e.\
with $N_\text{k} \ge 20$) are not a good match to SN~2005hk: though
only slightly brighter than SN~2005hk in $B$ band, their peak
magnitudes in the redder bands are significantly too bright compared
to SN~2005hk: the peak magnitudes in $R$ and $I$ reach values typical
of normal SNe~Ia.  However, the singly peaked $R$- and $I$-band light
curves of these models and their red $B - V$ colours at maximum light
are inconsistent with observations of normal SNe~Ia.  Since these
models would be sufficiently bright to be detected easily, the absence
of any such objects in the observed transient sample could indicate
that such vigorously ignited deflagrations are not realized in nature.
Alternatively, it could also indicate that in this case the flame is
more likely to undergo a DDT\@.  For the evolution of our model sample
assuming a DDT during the flame evolution, see
\citetalias{seitenzahl2013a} and \citet{sim2013a}.

Models N1def and N3def are significantly fainter than SN~2005hk.
N3def provides a decent match to the subluminous 1991bg-like SNe
\citep{filippenko1992b, leibundgut1993a} in bolometric and $B$-band
light curves [see comparison to the 1991bg-like SN~2005bl
\citep{taubenberger2008a} in Fig.~\ref{fig:lc}].  However, N3def
cannot explain the red colours typical for those subluminous events
(the model is too bright in $U$ band and too faint in \mbox{$V$-},
$R$- and $I$-band light curves compared to SN~2005bl).  Instead, the
colours and spectral properties (see next section) of models N1def and
N3def are qualitatively similar to those of the brighter model N5def.
Thus, N1def and N3def are promising candidates to explain some of the
fainter members of the class of SN~2002cx-like SNe (or SNe~Iax as
recently introduced by \citealt{foley2013b}), which currently consists
of 25 objects.

However, even N1def ($M_\text{max}^V=-16.84$~mag) is significantly
brighter than SN~2008ha ($M_\text{max}^V=-14.19$~mag) the faintest
object proposed to be a member of the class \citep{foley2009a,
  foley2013b}.  From our present model sample it seems impossible to
explain objects like SN~2008ha as the result of deflagrations in
Chandrasekhar-mass WDs.  Given the good agreement of N5def with
SN~2005hk this could indicate that different explosion mechanisms are
at work for different types of SNe~Iax.  However, we have not explored
all possible ignition configurations yet.  In particular, we have
neither varied the degree of off-set of a single ignition spot from
the centre in our model sample, nor the chemical composition of the
initial WD\@.  This will be the subject of a future dedicated study.

Note, however, that at least two of the SNe presented in
\citet{foley2013b} showed helium lines in their spectra.  This would
be difficult to explain in the canonical Chandrasekhar-mass
single-degenerate scenario, where a WD accretes hydrogen-rich matter
from a slightly evolved main-sequence or a red giant star.  However,
it is possible that a CO WD may reach the Chandrasekhar mass by
accretion from a helium-burning star \citep{iben1987a, ruiter2011a,
  wang2012b}, in which case one may expect to observe helium lines in
the spectra (but, most of the helium will stably be burnt to carbon).
On the other hand, more than one explosion mechanism might be at work
for SNe~Iax.  \citet{foley2009a, foley2010a}, for example, suggested
deflagrations of helium-accreting sub-Chandrasekhar-mass WDs, while
\citet{valenti2009a} and \citet{moriya2010a} favoured a core-collapse
origin.

\subsection{Spectra}
\label{sec:synth_obs_spectra}

\begin{figure}
  \centering
  \includegraphics[width=\columnwidth]{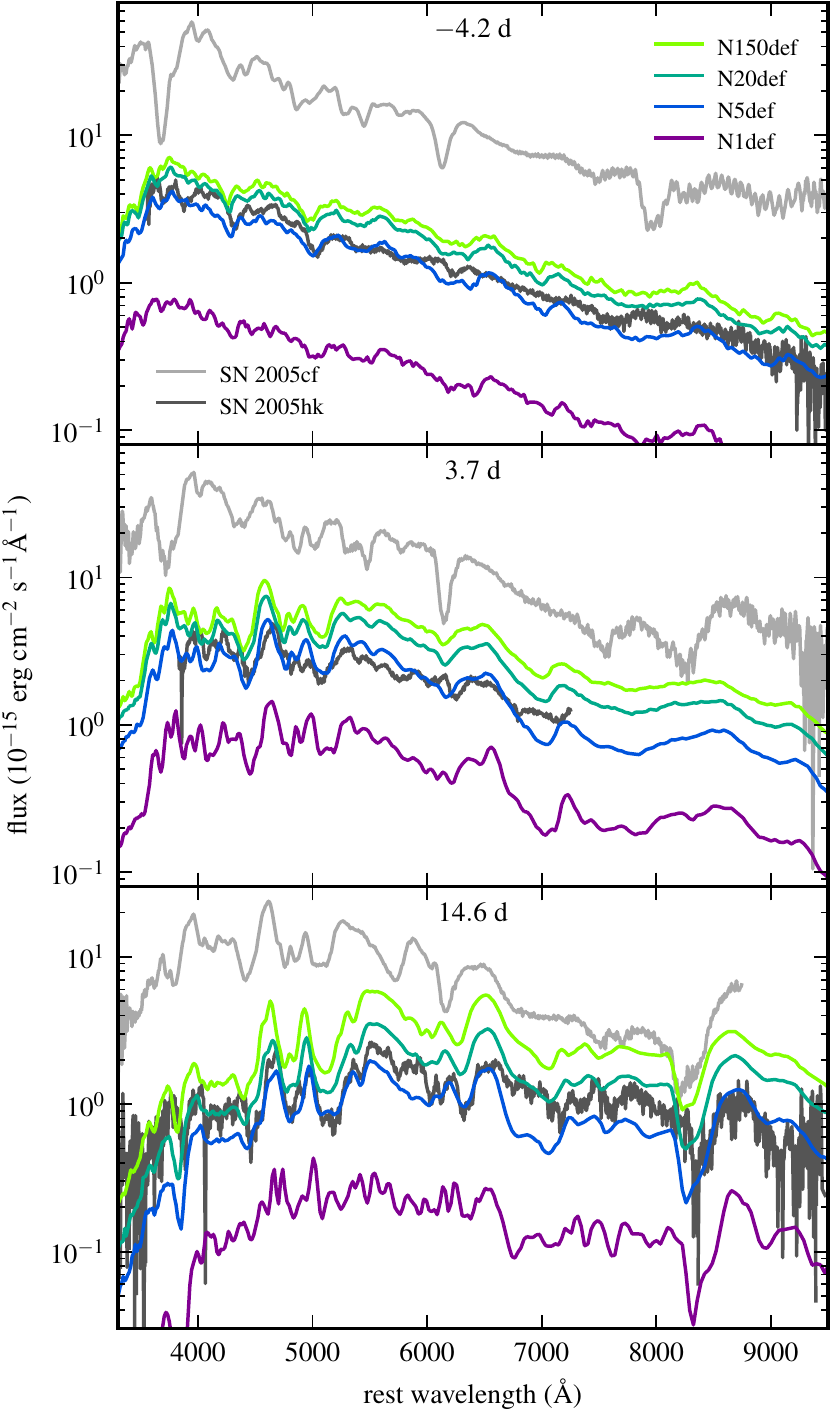}
  \caption{Spectral evolution for select models (colour coded) of our
    sample.  Shown are snapshots for $-4.2$, $3.7$ and $14.6$~d
    relative to $B$-band maximum (from top to bottom; see
    Table~\ref{tab:lc_props} for the rise times of the different
    models).  For comparison, observed spectra of SN~2005hk (dark
    grey; \citealt{phillips2007a}) and SN~2005cf (grey;
    \citealt{garavini2007a}) are overplotted for corresponding epochs.
    The observations were de-reddened and de-redshifted.}
  \label{fig:spectra}
\end{figure}

Fig.~\ref{fig:spectra}
shows angle-averaged synthetic spectra for select models that cover
the full distribution of ejecta and \nuc{56}{Ni} masses of our
simulations and are thus representative of the full sample.
Specifically, we focus on models N1def, N5def, N20def and N150def
spanning a range of $0.08$--$1.4~\msol$ in ejecta mass and
$0.03$--$0.38~\msol$ in \nuc{56}{Ni} mass (cf.\
Table~\ref{tab:energetics}).

Apart from differences in the absolute flux level, due to the
increasing \nuc{56}{Ni} mass for more vigorously ignited models, the
spectral shape along our model sequence is remarkably similar.  One
systematic difference is the increasing blueshift and broadening of
line features that reflects the increasing ejecta velocities along the
model sequence with increasing strength of the deflagration.

None of the deflagration models of our sample can account for the
observed spectra of normal SNe~Ia (e.g.\ SN~2005cf) since neither the
spectral features nor the absolute flux distribution match.  In
particular, our early time model spectra lack the strong absorption
features, associated with atomic lines of IMEs, such as Si, S and Ca,
that are characteristic of normal SNe~Ia.

Instead, the spectral features of our deflagration models provide a
good match to 2002cx-like SNe.  As discussed by \citet{kromer2013a},
in particular model N5def nicely reproduces the overall flux level and
spectral features of SN~2005hk, a prototypical 2002cx-like event.
Comparably good agreement is found for model N10def.  The more
vigorously ignited models such as N20def and N150def are too bright
compared to SN~2005hk, while the flux for models with a weaker
deflagration (e.g.\ N1def) is too low.

\subsection{Viewing-angle dependence}
\label{sec:synth_obs_los}

In the previous sections, we have discussed angle-averaged light curves
and spectra of our deflagration models.  However, since our models
show (more or less pronounced) large-scale asymmetries (see
Section~\ref{sec:vel_nucl}), the synthetic observables depend on the
viewing angle of the observer.  This is taken into account by our 3D
radiative transfer code \textsc{artis}.

\begin{figure*}
  \centering
  \includegraphics{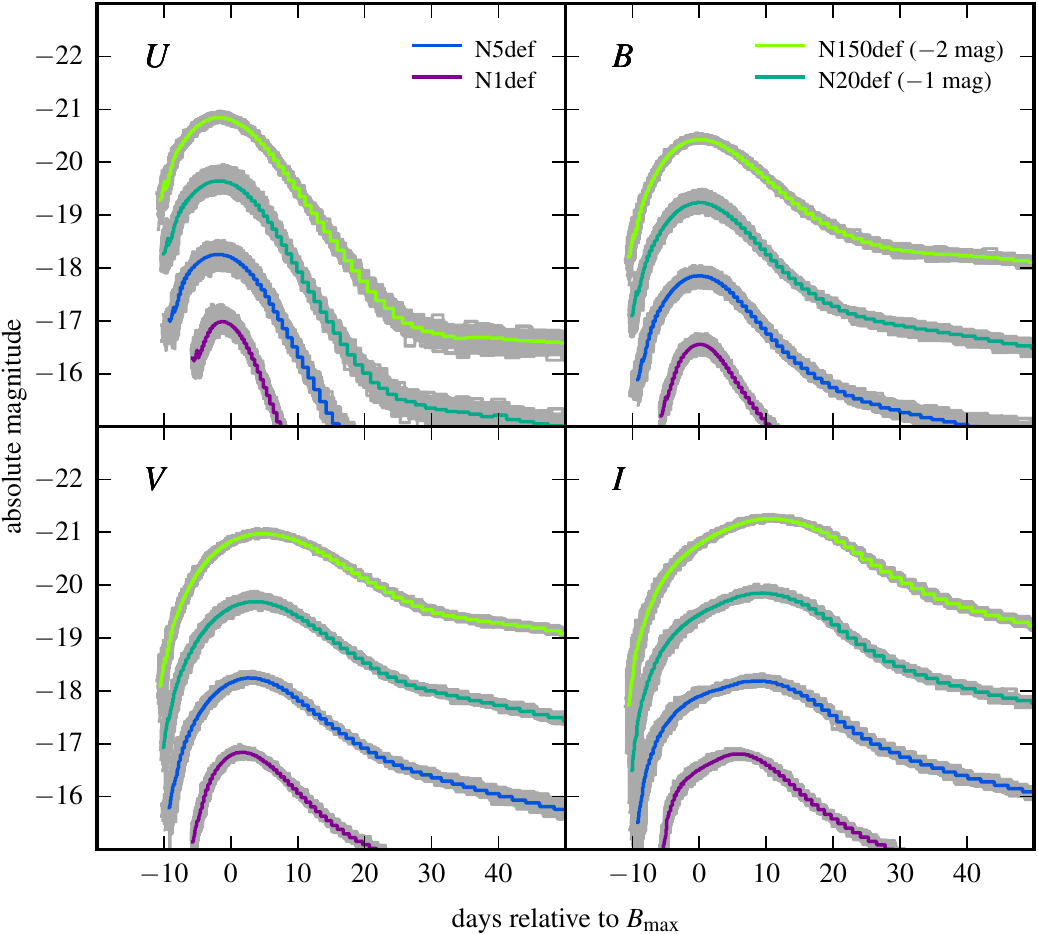}
  \caption{Synthetic light curves as seen from 100 different viewing
    angles (each of equal solid angle) for select models of our sample
    in grey.  The coloured lines show the angle-averaged light curves
    for comparison.  For clarity, the light curves of models N20def
    and N150def are shifted by $-1$ and $-2$~mag, respectively.}
  \label{fig:los_lcs}
\end{figure*}

Fig.~\ref{fig:los_lcs}
shows line-of-sight dependent synthetic light curves for four models
(N1def, N5def, N20def and N150def) that cover the full distribution of
ejecta and \nuc{56}{Ni} masses predicted by our set of simulations.
As for other explosion models \citep[see e.g.][]{kromer2009a,
  kromer2010a, roepke2012a}, the line-of-sight dependence weakens from
blue to red bands since the optical depth decreases with wavelength.

As discussed in Section~\ref{sec:vel_nucl}, weak deflagrations (N1def,
N5def and N20def) tend to evolve in a one-sided manner and preserve
parts of this asymmetry into the homologous expansion phase.  Thus,
these models tend to show a slightly stronger viewing-angle
sensitivity than vigorously ignited models with intermediate
$N_\text{k}$, such as N150def, that evolve in a very symmetric way and
show little large-scale asymmetry.  In general, however, the
viewing-angle dependence in all of our models is moderate.

\section{Summary}
\label{sec:disc}

To study the question of whether pure deflagrations of
Chan\-dra\-se\-khar-mass WDs in the single-degenerate scenario
contribute to the observed sample of SNe~Ia, we have carried out an
extensive study of explosion models.  This work presents 3D full-star
hydrodynamic simulations for a wide range of explosion strengths,
combined with detailed nucleosynthesis and 3D radiative transfer
calculations that provide synthetic observables, which can be
\emph{directly} compared to observations.  This goes beyond previous
studies of model sequences, which assess the validity of the pure
deflagration explosion mechanism based only on holistic, qualitative
arguments.

The major result of the hydrodynamic simulations is the occurrence of
a bound remnant in sparsely ignited deflagrations ($N_\text{k}
\lesssim 100$) with an energy release $E_\text{nuc} \lesssim 1.1
\times 10^{51}$~erg \citep[cf.][]{jordan2012b}.  The remnant is mostly
comprised of the unburnt matter that remains in the centre of the
star.  Most of the hot deflagration ashes are ejected and reach
homologous expansion.  We find a roughly linear relation between the
ejecta mass (and also the $^{56}$Ni mass) and $E_\text{nuc}$ (see
Fig.~\ref{fig:unbound_mass}).  The remnant masses in our sample lie
between $1.32$ and $0.09~\msol$ and the ejected $^{56}$Ni masses are
in the range $0.035$--$0.38$~\msol.  Pulsations in the bound material
were found to not re-ignite explosive burning above a certain
deflagration strength (for $N_\text{k} \ge 5$).  However, for our two
weakest deflagrations (N1def and N3def), re-ignition of the remnant
could not be excluded.

The bound remnants are enriched (mainly in the outer layers) with
3--10 per cent of IGE- and IME-rich deflagration ashes that were not
accelerated to escape velocity.  In our simulations performed on a
grid co-expanding with the ejecta, we find kick velocities of the
remnants of the order of $v_\text{kick} \le 36~\kms$ (both with our
monopole gravity solver as well as in two tests with a fast Fourier
transformation-based gravity solver).  Such low kick velocities are
contrary to the recent results of \citet{jordan2012b} and may be
insufficient for the ejection of the remnant objects from the binary
systems.

As the explosion energies of our models are low, the same is true also
for the maximum ejecta velocities, which lie in the range
$9000$--$14\,000~\kms$.  The ejecta show the complex structures
typical for deflagrations and both $^{56}$Ni and unburnt C/O material
can be found at all ejecta velocities.  Interestingly, the fraction of
the ejecta comprised of burnt material increases with decreasing
deflagration strength.

According to our synthetic colour light curves and spectra, the two
relatively sparsely ignited models N5def and N10def (which leave bound
remnants) are promising candidates for 2002cx-like SNe.  More
vigorously ignited models are neither comparable to 2002cx-like SNe
(too bright at red wavelengths), nor to normal SNe~Ia ($R$- and
$I$-band light curves are only singly peaked, $B - V$ colours at
maximum light disagree and the model spectra lack strong lines of IMEs
such as Si, S and Ca).  More sparsely ignited models may be
interesting candidates for some of the fainter members of 2002cx-like
SNe.  But, the faintest observed events (like SN~2008ha) of the
\citet{foley2013b} SN~Iax seem to be out of reach for our current set
of models.

A restricted range of sparse one-sided ignition configurations (N5def,
N10def) with an energy release $E_\text{nuc} \sim 0.5 \times
10^{51}$~erg leads to models that account for the observed properties
of SN~2002cx-like events.  Although we initiate our deflagrations in
the multiple ignition spot parametrization, this range of ignition
conditions bears some similarity with the findings of recent
pre-ignition simulations \citep{nonaka2012a} that predict off-centre
ignition in a single bubble at a radial distance in the range
$40$--$75$~km.  We find that more vigorously ignited deflagrations
($N_\text{k} \gtrsim 20$) do not resemble any observed class of SN~Ia.
This could either mean that such ignition conditions are not realized
in nature.  Alternatively, perhaps strong deflagrations always trigger
a DDT, leading to brighter events: our DDT models with $N_\text{k}
\sim 100$ have been shown to compare well with normal SNe~Ia
(\citealt{roepke2012a}; \citetalias{seitenzahl2013a};
\citealt{sim2013a}).

\section*{Acknowledgements}
The simulations presented here were carried out in part on the JUGENE
supercomputer at the Forschungszentrum J{\"u}lich within the
Partnership for Advanced Computing in Europe (PRA026), the grant HMU13
and in part at the Computer Center of the Max Planck Society,
Garching, Germany.  This work was also supported by the Deutsche
Forschungsgemeinschaft via the Transregional Collaborative Research
Center TRR 33 `The Dark Universe', the Emmy Noether Programme (RO
3676/1-1), the ARCHES prize of the German Ministry of Education and
Research (BMBF), the graduate school `Theoretical Astrophysics and
Particle Physics' at the University of W\"urzburg (GRK 1147) and the
Excellence Cluster EXC~153.  MF, FKR and SAS acknowledge travel
support by the DAAD/Go8 German--Australian exchange programme.  RP
acknowledges support by the European Research Council (ERC-StG grant
EXAGAL-308037).

{\footnotesize%

}

\appendix

\section{Deflagration tables}
\label{sec:deftab}

\begin{figure}
  \centering
  \includegraphics[width=\columnwidth]{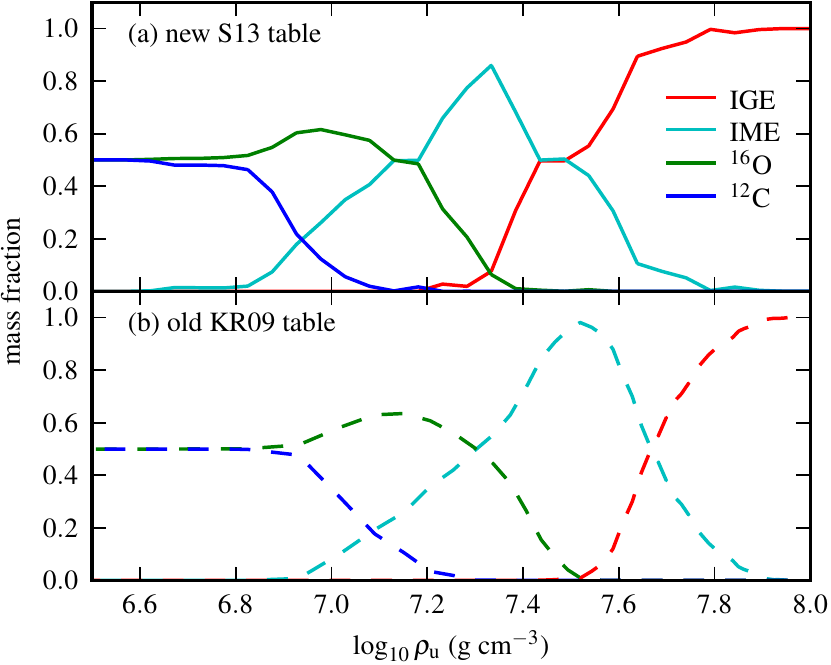}
  \caption{Mass fractions of the reduced species in our CO
    deflagration tables against the density of the unburnt fuel
    $\rho_\text{u}$.}
  \label{fig:deftab}
\end{figure}

\begin{figure}
  \centering
  \includegraphics[width=\columnwidth]{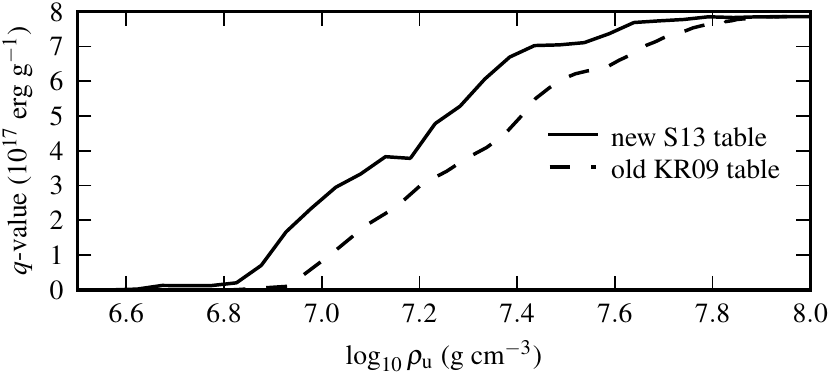}
  \caption{Reaction $q$-values against $\rho_\text{u}$ for our CO
    deflagration tables.}
  \label{fig:deftab_q}
\end{figure}

To calibrate the energy release in the deflagration, we have used an
iterative scheme in which we alternate between a sample hydrodynamic
explosion simulation and detailed nucleosynthesis post-processing
calculations, as described in the appendix of \citet{fink2010a}.  We
have tabulated the final composition of deflagration ashes in CO
matter for our reduced set of species (see Section~\ref{sec:hd}) as a
function of the density in the unburnt matter, $\rho_\text{u}$.  In
contrast to previous studies, in which we used our old KR09
deflagration table (\citealt*{kasen2009a}; \citealt{seitenzahl2010a};
\citealt*{seitenzahl2011a}; \citealt{travaglio2011a};
\citealt*{ciaraldi2013a}; \citealt{parikh2013a}; see
Fig.~\ref{fig:deftab}b),
here, we use an improved algorithm to determine the value of
$\rho_\text{u}$ in `mixed' cells intersected by the flame (the old
algorithm overestimated $\rho_\text{u}$ in some cases).  Our new S13
deflagration table is shown in Fig.~\ref{fig:deftab}(a) and is used
in this study and in \citet{seitenzahl2013a}.  The differences between
the old and the new table are moderate: as shown in
Fig.~\ref{fig:deftab_q},
the reaction $q$-value curve is shifted somewhat towards lower fuel
densities $\rho_\text{u}$.  With the new table, the consistency
between the nuclear energy release in our hydrodynamic simulations and
our detailed post-processing nucleosynthesis results is improved.

\section{Detailed nucleosynthetic yields}
\label{sec:nucl_detailed}

Total integrated nucleosynthetic yields of stable and radioactive
nuclei in the model ejecta are given in Tables~\ref{tab:syields} and
\ref{tab:ryields}.  Tables~\ref{tab:syields_core} and
\ref{tab:ryields_core} provide the respective data for the remnant
objects.  For radioactive nuclei, we simply tabulate the values at the
end of our post-processing calculation ($t = 100$~s).  To determine
the yields of stable nuclei, we decay all radioactive nuclei with
half-lives of less than $2$~Gyr.  Yields of radioactive nuclei with
longer half-lives are given with their $t = 100$~s value among the
stable isotopes.

\begin{table*}
  \centering
  \begin{sideways}
    \begin{minipage}{680pt}
      \caption{Asymptotic yields of stable isotopes in the model
        ejecta in solar masses.}
      \label{tab:syields}
\begin{tabular}{@{}*{14}{l@{\hspace{1.2ex}}r@{$\,\times\,$}}l@{}}
        \hline
                  & \multicolumn{2}{c}{N1def} & \multicolumn{2}{c}{N3def} & \multicolumn{2}{c}{N5def} & \multicolumn{2}{c}{N10def} & \multicolumn{2}{c}{N20def} & \multicolumn{2}{c}{N40def} & \multicolumn{2}{c}{N100Ldef} & \multicolumn{2}{c}{N100def} & \multicolumn{2}{c}{N100Hdef} & \multicolumn{2}{c}{N150def} & \multicolumn{2}{c}{N200def} & \multicolumn{2}{c}{N300Cdef} & \multicolumn{2}{c}{N1600def} & \multicolumn{2}{c}{N1600Cdef} \\
        \hline
        $M_\text{tot,ej}$
                  & $8.43$ & $10^{-2}$  & $1.95$ & $10^{-1}$  & $3.72$ & $10^{-1}$  & $4.78$ & $10^{-1}$  & $8.59$ & $10^{-1}$  & $1.21$ & $10^{+0}$  & $1.23$ & $10^{+0}$  & $1.31$ & $10^{+0}$  & $1.31$ & $10^{+0}$  & $1.40$ & $10^{+0}$  & $1.40$ & $10^{+0}$  & $1.40$ & $10^{+0}$  & $1.40$ & $10^{+0}$  & $1.40$ & $10^{+0}$  \\
\end{tabular}
    \end{minipage}
  \end{sideways}
\end{table*}
\begin{table*}
  \centering
  \begin{sideways}
    \begin{minipage}{680pt}
      \contcaption{}
      \begin{tabular}{@{}*{14}{l@{\hspace{1.2ex}}r@{$\,\times\,$}}l@{}}
        \hline
                  & \multicolumn{2}{c}{N1def} & \multicolumn{2}{c}{N3def} & \multicolumn{2}{c}{N5def} & \multicolumn{2}{c}{N10def} & \multicolumn{2}{c}{N20def} & \multicolumn{2}{c}{N40def} & \multicolumn{2}{c}{N100Ldef} & \multicolumn{2}{c}{N100def} & \multicolumn{2}{c}{N100Hdef} & \multicolumn{2}{c}{N150def} & \multicolumn{2}{c}{N200def} & \multicolumn{2}{c}{N300Cdef} & \multicolumn{2}{c}{N1600def} & \multicolumn{2}{c}{N1600Cdef} \\
        \hline
        $^{44}$Ca & $1.97$ & $10^{-7}$  & $4.83$ & $10^{-7}$  & $8.98$ & $10^{-7}$  & $1.12$ & $10^{-6}$  & $1.90$ & $10^{-6}$  & $2.16$ & $10^{-6}$  & $2.07$ & $10^{-6}$  & $2.30$ & $10^{-6}$  & $2.58$ & $10^{-6}$  & $2.52$ & $10^{-6}$  & $2.33$ & $10^{-6}$  & $2.46$ & $10^{-6}$  & $2.10$ & $10^{-6}$  & $2.15$ & $10^{-6}$  \\
        $^{46}$Ca & $4.82$ & $10^{-13}$ & $1.22$ & $10^{-12}$ & $1.97$ & $10^{-12}$ & $2.70$ & $10^{-12}$ & $6.42$ & $10^{-12}$ & $6.42$ & $10^{-12}$ & $5.59$ & $10^{-12}$ & $6.77$ & $10^{-12}$ & $6.67$ & $10^{-10}$ & $7.26$ & $10^{-12}$ & $6.12$ & $10^{-12}$ & $1.06$ & $10^{-11}$ & $5.47$ & $10^{-12}$ & $5.75$ & $10^{-12}$ \\
        $^{48}$Ca & $7.17$ & $10^{-16}$ & $1.92$ & $10^{-15}$ & $5.29$ & $10^{-15}$ & $5.09$ & $10^{-15}$ & $1.51$ & $10^{-14}$ & $3.07$ & $10^{-15}$ & $1.03$ & $10^{-16}$ & $5.12$ & $10^{-15}$ & $9.58$ & $10^{-11}$ & $2.15$ & $10^{-15}$ & $3.03$ & $10^{-16}$ & $8.17$ & $10^{-14}$ & $9.48$ & $10^{-17}$ & $1.06$ & $10^{-16}$ \\
        $^{45}$Sc & $3.03$ & $10^{-9}$  & $8.45$ & $10^{-9}$  & $1.40$ & $10^{-8}$  & $1.89$ & $10^{-8}$  & $3.88$ & $10^{-8}$  & $4.05$ & $10^{-8}$  & $3.50$ & $10^{-8}$  & $4.15$ & $10^{-8}$  & $5.35$ & $10^{-8}$  & $4.61$ & $10^{-8}$  & $4.17$ & $10^{-8}$  & $4.60$ & $10^{-8}$  & $3.70$ & $10^{-8}$  & $3.71$ & $10^{-8}$  \\
\hline
      \end{tabular}
    \end{minipage}
  \end{sideways}
\end{table*}
\begin{table*}
  \centering
  \begin{sideways}
    \begin{minipage}{680pt}
      \caption{Yields of radioactive isotopes in the model ejecta in
        solar masses at $t = 100$~s.}
      \label{tab:ryields}
      \begin{tabular}{@{}*{14}{l@{\hspace{1.2ex}}r@{$\,\times\,$}}l@{}}
        \hline
                  & \multicolumn{2}{c}{N1def} & \multicolumn{2}{c}{N3def} & \multicolumn{2}{c}{N5def} & \multicolumn{2}{c}{N10def} & \multicolumn{2}{c}{N20def} & \multicolumn{2}{c}{N40def} & \multicolumn{2}{c}{N100Ldef} & \multicolumn{2}{c}{N100def} & \multicolumn{2}{c}{N100Hdef} & \multicolumn{2}{c}{N150def} & \multicolumn{2}{c}{N200def} & \multicolumn{2}{c}{N300Cdef} & \multicolumn{2}{c}{N1600def} & \multicolumn{2}{c}{N1600Cdef} \\
        \hline
        $M_\text{tot,ej}$
                  & $8.43$ & $10^{-2}$  & $1.95$ & $10^{-1}$  & $3.72$ & $10^{-1}$  & $4.78$ & $10^{-1}$  & $8.59$ & $10^{-1}$  & $1.21$ & $10^{+0}$  & $1.23$ & $10^{+0}$  & $1.31$ & $10^{+0}$  & $1.31$ & $10^{+0}$  & $1.40$ & $10^{+0}$  & $1.40$ & $10^{+0}$  & $1.40$ & $10^{+0}$  & $1.40$ & $10^{+0}$  & $1.40$ & $10^{+0}$  \\
        $^{14}$C  & $1.24$ & $10^{-6}$  & $3.94$ & $10^{-6}$  & $4.85$ & $10^{-6}$  & $7.21$ & $10^{-6}$  & $2.52$ & $10^{-5}$  & $4.58$ & $10^{-5}$  & $2.68$ & $10^{-5}$  & $5.52$ & $10^{-5}$  & $4.93$ & $10^{-5}$  & $5.31$ & $10^{-5}$  & $4.07$ & $10^{-5}$  & $5.79$ & $10^{-5}$  & $3.74$ & $10^{-5}$  & $2.88$ & $10^{-5}$  \\
        $^{22}$Na & $1.10$ & $10^{-9}$  & $3.48$ & $10^{-9}$  & $4.64$ & $10^{-9}$  & $8.06$ & $10^{-9}$  & $4.56$ & $10^{-8}$  & $1.21$ & $10^{-7}$  & $2.32$ & $10^{-8}$  & $9.57$ & $10^{-8}$  & $1.93$ & $10^{-7}$  & $7.30$ & $10^{-8}$  & $3.49$ & $10^{-8}$  & $6.21$ & $10^{-8}$  & $2.64$ & $10^{-8}$  & $2.36$ & $10^{-8}$  \\
        $^{26}$Al & $1.26$ & $10^{-7}$  & $4.09$ & $10^{-7}$  & $5.23$ & $10^{-7}$  & $8.93$ & $10^{-7}$  & $4.84$ & $10^{-6}$  & $1.22$ & $10^{-5}$  & $2.78$ & $10^{-6}$  & $1.06$ & $10^{-5}$  & $1.67$ & $10^{-5}$  & $8.29$ & $10^{-6}$  & $4.30$ & $10^{-6}$  & $7.52$ & $10^{-6}$  & $3.27$ & $10^{-6}$  & $2.82$ & $10^{-6}$  \\
\hline
      \end{tabular}
    \end{minipage}
  \end{sideways}
\end{table*}

\begin{table*}
  \centering
  \begin{minipage}{493pt}
    \caption{Asymptotic yields of stable isotopes in the bound
      remnants in solar masses.}
    \label{tab:syields_core}
    \begin{tabular}{@{}*{9}{lr@{$\,\times\,$}}l@{}}
      \hline
                & \multicolumn{2}{c}{N1def} & \multicolumn{2}{c}{N3def} & \multicolumn{2}{c}{N5def} & \multicolumn{2}{c}{N10def} & \multicolumn{2}{c}{N20def} & \multicolumn{2}{c}{N40def} & \multicolumn{2}{c}{N100Ldef} & \multicolumn{2}{c}{N100def} & \multicolumn{2}{c}{N100Hdef} \\
      \hline
      $M_\text{tot,b}$
                & $1.32$ & $10^{+0}$  & $1.21$ & $10^{+0}$  & $1.03$ & $10^{+0}$  & $9.26$ & $10^{-1}$  & $5.45$ & $10^{-1}$  & $1.90$ & $10^{-1}$  & $1.33$ & $10^{-1}$  & $9.0$  & $10^{-2}$  & $1.02$ & $10^{-1}$  \\
      $^{4}$He  & $2.85$ & $10^{-5}$  & $9.64$ & $10^{-6}$  & $1.54$ & $10^{-5}$  & $1.66$ & $10^{-5}$  & $5.73$ & $10^{-6}$  & $1.14$ & $10^{-6}$  & $4.79$ & $10^{-6}$  & $5.71$ & $10^{-7}$  & $4.50$ & $10^{-8}$  \\
\end{tabular}
  \end{minipage}
\end{table*}
\begin{table*}
  \centering
  \begin{minipage}{493pt}
    \contcaption{}
    \begin{tabular}{@{}*{9}{lr@{$\,\times\,$}}l@{}}
      \hline
                & \multicolumn{2}{c}{N1def} & \multicolumn{2}{c}{N3def} & \multicolumn{2}{c}{N5def} & \multicolumn{2}{c}{N10def} & \multicolumn{2}{c}{N20def} & \multicolumn{2}{c}{N40def} & \multicolumn{2}{c}{N100Ldef} & \multicolumn{2}{c}{N100def} & \multicolumn{2}{c}{N100Hdef} \\
      \hline
      $^{63}$Cu & $1.84$ & $10^{-8}$  & $1.01$ & $10^{-8}$  & $1.28$ & $10^{-8}$  & $1.18$ & $10^{-8}$  & $3.78$ & $10^{-9}$  & $1.28$ & $10^{-9}$  & $1.85$ & $10^{-9}$  & $3.31$ & $10^{-10}$ & $2.89$ & $10^{-9}$  \\
      $^{65}$Cu & $8.13$ & $10^{-9}$  & $2.12$ & $10^{-9}$  & $2.94$ & $10^{-9}$  & $2.82$ & $10^{-9}$  & $9.83$ & $10^{-10}$ & $2.15$ & $10^{-10}$ & $7.57$ & $10^{-10}$ & $6.17$ & $10^{-11}$ & $3.53$ & $10^{-10}$ \\
      $^{64}$Zn & $9.71$ & $10^{-8}$  & $3.30$ & $10^{-8}$  & $4.57$ & $10^{-8}$  & $4.20$ & $10^{-8}$  & $1.44$ & $10^{-8}$  & $3.00$ & $10^{-9}$  & $1.10$ & $10^{-8}$  & $9.82$ & $10^{-10}$ & $1.95$ & $10^{-10}$ \\
      $^{66}$Zn & $1.37$ & $10^{-7}$  & $4.44$ & $10^{-8}$  & $6.36$ & $10^{-8}$  & $6.02$ & $10^{-8}$  & $2.11$ & $10^{-8}$  & $4.32$ & $10^{-9}$  & $1.80$ & $10^{-8}$  & $1.45$ & $10^{-9}$  & $2.45$ & $10^{-10}$ \\
      $^{67}$Zn & $1.14$ & $10^{-10}$ & $3.24$ & $10^{-11}$ & $4.42$ & $10^{-11}$ & $3.97$ & $10^{-11}$ & $1.38$ & $10^{-11}$ & $2.82$ & $10^{-12}$ & $1.11$ & $10^{-11}$ & $8.53$ & $10^{-13}$ & $1.59$ & $10^{-12}$ \\
      $^{68}$Zn & $5.08$ & $10^{-11}$ & $2.08$ & $10^{-11}$ & $2.96$ & $10^{-11}$ & $2.79$ & $10^{-11}$ & $9.69$ & $10^{-12}$ & $2.49$ & $10^{-12}$ & $6.63$ & $10^{-12}$ & $6.37$ & $10^{-13}$ & $1.82$ & $10^{-11}$ \\
      $^{70}$Zn & $7.57$ & $10^{-17}$ & $1.12$ & $10^{-16}$ & $1.70$ & $10^{-16}$ & $2.31$ & $10^{-16}$ & $7.43$ & $10^{-17}$ & $2.04$ & $10^{-17}$ & $8.31$ & $10^{-29}$ & $4.83$ & $10^{-18}$ & $5.73$ & $10^{-15}$ \\
\hline
    \end{tabular}
  \end{minipage}
\end{table*}
\begin{table*}
  \centering
  \begin{minipage}{493pt}
    \caption{Yields of radioactive isotopes in the bound remnants in
      solar masses at $t = 100$~s.}
    \label{tab:ryields_core}
    \begin{tabular}{@{}*{9}{lr@{$\,\times\,$}}l@{}}
      \hline
                & \multicolumn{2}{c}{N1def} & \multicolumn{2}{c}{N3def} & \multicolumn{2}{c}{N5def} & \multicolumn{2}{c}{N10def} & \multicolumn{2}{c}{N20def} & \multicolumn{2}{c}{N40def} & \multicolumn{2}{c}{N100Ldef} & \multicolumn{2}{c}{N100def} & \multicolumn{2}{c}{N100Hdef} \\
      \hline
      $M_\text{tot,b}$
                & $1.32$ & $10^{+0}$  & $1.21$ & $10^{+0}$  & $1.03$ & $10^{+0}$  & $9.26$ & $10^{-1}$  & $5.45$ & $10^{-1}$  & $1.90$ & $10^{-1}$  & $1.33$ & $10^{-1}$  & $9.0$  & $10^{-2}$  & $1.02$ & $10^{-1}$ \\
      $^{14}$C  & $2.41$ & $10^{-4}$  & $1.81$ & $10^{-4}$  & $1.22$ & $10^{-4}$  & $1.17$ & $10^{-4}$  & $8.67$ & $10^{-5}$  & $3.55$ & $10^{-5}$  & $1.51$ & $10^{-6}$  & $1.04$ & $10^{-5}$  & $1.72$ & $10^{-5}$  \\
      $^{22}$Na & $2.33$ & $10^{-7}$  & $2.08$ & $10^{-7}$  & $1.84$ & $10^{-7}$  & $1.77$ & $10^{-7}$  & $1.31$ & $10^{-7}$  & $6.23$ & $10^{-8}$  & $1.16$ & $10^{-9}$  & $1.19$ & $10^{-8}$  & $8.84$ & $10^{-8}$  \\
      $^{26}$Al & $3.29$ & $10^{-5}$  & $2.86$ & $10^{-5}$  & $2.42$ & $10^{-5}$  & $2.33$ & $10^{-5}$  & $1.77$ & $10^{-5}$  & $8.72$ & $10^{-6}$  & $1.41$ & $10^{-7}$  & $1.44$ & $10^{-6}$  & $7.74$ & $10^{-6}$  \\
      $^{32}$Si & $1.15$ & $10^{-7}$  & $1.19$ & $10^{-7}$  & $6.92$ & $10^{-8}$  & $6.69$ & $10^{-8}$  & $4.50$ & $10^{-8}$  & $1.70$ & $10^{-8}$  & $1.55$ & $10^{-9}$  & $3.29$ & $10^{-9}$  & $8.68$ & $10^{-9}$  \\
      $^{32}$P  & $5.28$ & $10^{-7}$  & $6.33$ & $10^{-7}$  & $3.92$ & $10^{-7}$  & $4.06$ & $10^{-7}$  & $2.38$ & $10^{-7}$  & $7.16$ & $10^{-8}$  & $1.82$ & $10^{-8}$  & $1.73$ & $10^{-8}$  & $1.55$ & $10^{-7}$  \\
      $^{33}$P  & $1.25$ & $10^{-7}$  & $1.58$ & $10^{-7}$  & $1.06$ & $10^{-7}$  & $1.16$ & $10^{-7}$  & $5.74$ & $10^{-8}$  & $1.57$ & $10^{-8}$  & $9.00$ & $10^{-9}$  & $5.60$ & $10^{-9}$  & $2.65$ & $10^{-8}$  \\
\hline
    \end{tabular}

    \bigskip

    \footnotesize{This paper has been typeset from a \TeX / \LaTeX\
      file prepared by the author.}
  \end{minipage}
\end{table*}


\end{document}